\documentclass[11pt]{revtex4-2}

\usepackage{hyperref}
\usepackage{amsmath,amsfonts,amssymb}
\hypersetup{dvips,dvipdfm,colorlinks=true,urlcolor=magenta,filecolor=magenta,linktoc=page,citecolor=red,linkcolor=blue,bookmarks=true}
\usepackage{graphicx,epstopdf}
\usepackage{multirow}

\begin{document}
	\title{The Raychaudhuri Equation in inhomogeneous FLRW space-time : A $f(R)$-gravity model}
	\author{Madhukrishna Chakraborty \footnote{chakmadhu1997@gmail.com}}
	\author{Akash Bose\footnote{bose.akash13@gmail.com}}
	\author{Subenoy Chakraborty\footnote{schakraborty.math@gmail.com}}
	\affiliation{Department of Mathematics, Jadavpur University, Kol - 700032, India}
	
	\begin{abstract}
		In general description of the Raychaudhuri equation it is found that this first order non-linear differential equation can be written as a second order linear differential equation in the form of Harmonic Oscillator with varying frequency. Further, the integrability of the Raychaudhuri equation has been studied and also the expansion scalar is obtained in an explicit form. Subsequently, $f(R)$ gravity theory has been studied in the background of inhomogeneous FLRW spacetime with an aim to formulate the Raychaudhuri Equation. A congruence of time-like geodesics has been investigated using the Raychaudhuri Equation to examine whether the geodesics converge or not and some possible conditions are determined to avoid singularity. Finally, a brief quantum description has been presented.
	\end{abstract}
 \maketitle

\section{Introduction}
The cosmologists have been facing a challenge for more than the last two decades to accommodate the observed accelerated expansion in the theoretical framework of gravity theories. One of the two possible ways towards finding a resolution of the accelerated expansion is to choose an alternative theory of gravity(other than Einstein gravity) without introducing any exotic matter i.e dark energy (DE) \cite{Capozziello:2019cav}. A common and natural generalization of Einstein gravity is obtained by replacing the Ricci scalar `$R$' in the Einstein-Hilbert action by an arbitrary function $f(R)$ \cite{Carroll:2003wy,Nojiri:2007as,Cognola:2007zu,Elizalde:2010ts,Nojiri:2010wj,Sotiriou:2008rp}. Also there are other well-known generalizations of Einstein gravity namely a non-minimally coupled scalar field theory \cite{Cid:2017wtf,Fontana:2018fof}, $f(T)$ gravity theory \cite{Bose:2020xdz,Cai:2015emx} etc. The Raychaudhuri equation (RE) has an immense contribution in modern cosmology and is a fundamental tool to study exact solutions of Einstein's equations in general relativity. \cite{Dunsby:2015ers}.
On the other hand, after the detection of gravitational waves, general relativity (GR) \cite{Wald:1984rg,Weinberg:1972kfs} is a universally accepted theory of gravity, despite the inherent existence of singularity in it as predicted by the famous singularity theorems of Hawking and Penrose \cite{Hawking:1973uf,Hawking:1970zqf,Penrose:1964wq}. The RE \cite{Raychaudhuri:1953yv}-\cite{Dadhich:2005qr}is the main ingredient behind these singularity theorems. The RE as it stands is purely a geometric identity in Riemannian geometry. However, it becomes a physical equation showing an equivalence between geometry and matter when gravity theory is imposed to determine the Ricci Tensor. In Einstein gravity, if the matter field satisfies the strong energy condition (SEC) then according to the RE, an initially converging congruence of time-like geodesics focuses within finite affine parameter value \cite{Poisson:2009pwt,Wald:1984}, leading to the formation of a congruence singularity (may or may not be a curvature singularity). The idea of Focusing theorem from RE together with some conditions on space-time geometry leads to the existence of singularity: The Singularity theorems \cite{Hawking:1970zqf,Penrose:1964wq}. In this context, one may say that the SEC or equivalent condition on Ricci tensor is termed as Convergence condition (CC) in Einstein gravity. This CC essentially indicates the attractive nature of gravity theory and leads to geodesic focusing. In the present work space-time is chosen as inhomogeneous Friedmann–Lemaître–Robertson–Walker (FLRW) model and $f(R)$ gravity theory \cite{Sotiriou:2008rp} has been constructed in this space-time. RE has been formulated for this model with the above $f(R)$ gravity theory. Finally, a congruence of time-like geodesics has been studied from the point of view of RE. The paper is organized in the following manner: Section II deals with a general description of the RE.
 In Section III, RE has been derived for $f(R)$-gravity in  an inhomogeneous FLRW model and its nature has been studied.
Also CC and scope of avoidance of singularity have been discussed.
In Section IV Geodesic congruences have been investigated from RE point of view. A quantization scheme has been discussed in section V .The paper ends with brief discussion and conclusion in section VI.
\section{Raychaudhuri Equation : A general description}
 The RE \cite{Kar:2008zz} essentially characterizes the kinematics of flows in a geometrical space. Usually, flows are generated by a vector field and in turn integral curves of the vector field identify the flow. This congruence of integral curves may be geodesic or non-geodesic in nature. In the Riemannian space these congruences are either time-like or null in nature. The REs are the evolution equations of the kinematic quantities which characterize the flow. Historically, only the evolution of the expansion scalar is the RE while the evolution of the other kinematic quantities are termed as Codazzi-Raychaudhuri equations.

Let `$l$' be the parameter identifying points on the integral curves and ${v^{a}}$ be the velocity vector field along those curves.Then gradient of the velocity vector field,a second rank tensor can be written as 
\begin{equation} B_{ab}=\nabla_{b}v_{a}=\sigma_{ab}+\omega_{ab}+\frac{1}{n-1}q_{ab}\Theta,
	\end{equation}
 where
  $$\sigma_{ab}=\sigma_{(ab)},~~\sigma^{a}_{a}=0,$$ i.e symmetric and traceless part of $B_{ab}$ having expression 
\begin{equation}\label{eq2*}
	\sigma_{(ab)}=\frac{1}{2}(\nabla_{a}v_{b}+\nabla_{b}v_{a})-\frac{1}{n-1}q_{ab}\Theta+\frac{1}{2}(A_{b}v_{a}+A_{a}v_{b}),
\end{equation}
 and is known as shear tensor.
 $\Theta=\nabla_{a}v^{a}$ is termed as expansion scalar and the antisymmetric rotation tensor $\omega_{ab}=\omega_{[ab]}$ has the expression \begin{equation}
 \omega_{ab}=\frac{1}{2}(\nabla_{b}v_{a}-\nabla_{a}v_{b})-\frac{1}{2}(A_{b}v_{a}-A_{a}v_{b}),
\end{equation}
where $A_{a}$is the acceleration vector, $n$ is the dimension of the space-time and $q_{ab}=g_{ab}+v_{a}v_{b}$ is the projection tensor with $v_{a}v^{a}=-1$ for time-like curves. Then the RE, the evolution equation for expansion scalar along the flow representing a time-like congruence is given by 
\begin{equation}\label{eq4*}
	\dfrac{\mathrm{d}\Theta}{\mathrm{d}l}=-\frac{1}{3}\Theta^{2}-\sigma_{ab}\sigma^{ab}+\omega_{ab}\omega^{ab}+\nabla_{b}A^{b}-R_{ab}v^{a}v^{b},
\end{equation}
 with $A^{a}=v^{b}\nabla_{b}v^{a}$, the acceleration vector field. The RE can be simplified if we assume i)the congruence of time-like geodesics (then the acceleration vector i.e $A^{b}=0$) and ii)the congruence is chosen to be hypersurface orthogonal (by Frobenius theorem $w_{ab}=0$). As a result, the simplified RE takes the form
  \begin{equation}
 	\dfrac{\mathrm{d}\Theta}{\mathrm{d}l}=-\dfrac{1}{3}\Theta^{2}-\sigma_{ab}\sigma^{ab}-R_{ab}v^{a}v^{b}.
 \end{equation}

 Note that no gravity theory has been used so far and the above RE is purely a geometric equation (identity). However, in Einstein gravity the last term on the r.h.s can be written as 
\begin{equation}
	R_{ab}v^{a}v^{b}=(T_{ab}-\frac{1}{2}Tg_{ab})v^{a}v^{b}.
\end{equation}

 Now, if the matter field satisfies the SEC (normal/usual matter) i.e 
 \begin{equation}
 	T_{ab}v^{a}v^{b}+\frac{1}{2}T \geqslant0,
 	\end{equation}
 then the RE gives \begin{equation}
 	\dfrac{\mathrm{d}\Theta}{\mathrm{d}l}+\dfrac{1}{3}\Theta^{2} \leqslant0.
 	\end{equation}
 
 This implies that an initially converging time-like geodesic congruence develops a caustic (i.e $\Theta\rightarrow-\infty$) within finite proper time (choosing the parameter to be proportional to the proper time). This is known as the Focusing theorem and the condition $R_{ab}v^{a}v^{b}\geqslant0$ is termed as CC for time-like geodesics. One can consider  $\Theta$, the expansion scalar as the rate of change of volume of the transverse subspace of the congruence. So $\Theta\rightarrow-\infty$ implies a convergence of the congruence while $\Theta\rightarrow+\infty$ indicates a total divergence. The RE, a first order non-linear equation, is of central importance in the context of the \textit{Singularity Theorems} \cite{Hawking:1970zqf,Penrose:1964wq}. Further, mathematically the RE is known as a Riccati equation, and it becomes a second order linear equation in the form of a harmonic oscillator equation with varying frequency as \cite{Horwitz:2021lyc,Kar:2006ms} \begin{equation}
 	\dfrac{\mathrm{d}^{2}X}{\mathrm{d}l^{2}}+\dfrac{1}{3}(R_{ab}v^{a}v^{b}+\sigma^{2}-\omega^{2})X=0,$$ with $$\Theta=\dfrac{3}{X}\dfrac{\mathrm{d}X}{\mathrm{d}l}=3\dfrac{\mathrm{d}(\ln X)}{\mathrm{d}l}.
 	\end{equation}
 
Now the above convergence condition can be stated as follows : \\i)$X$ is negative initially,\\ ii)$X=0$ at a finite value of the parameter to have a negatively infinite expansion. 

As $\Theta$ may be identified as the derivative of the geometric entropy (S) so one may identify S as $\ln X$. Here $X$ can be chosen as an average or effective geodesic deviation. Using the well-known Sturm Comparison theorem in the theory of differential equations, the criterion for the existence of zeros in $X$ at finite value of the affine parameter is given by 
\begin{equation}
	R_{ab}v^{a}v^{b}+\sigma^{2}-\omega^{2}\geqslant0.
\end{equation}

The above inequality for convergence of geodesic congruence shows that shear is in favour of convergence while rotation opposes the convergence. Hence for hyper-surface orthogonal congruence of geodesics the convergence condition is $R_{ab}v^{a}v^{b}>0$ (as given earlier). 

We shall now transform the RE suitably with an aim to study its integrability. For this we consider a congruence of time-like geodesics which are hyper-surface orthogonal. Let us define 
\begin{equation}
	\Lambda=\sqrt{\det q_{ab}}=\sqrt{q},
\end{equation}
 then the dynamical evolution of $q$ is given by \cite{Poisson:2009pwt} \begin{equation}
 	\dfrac{1}{\sqrt{q}}\dfrac{\mathrm{d}\sqrt{q}}{\mathrm{d}l}=\Theta,
 \end{equation} and hence \begin{equation}
 \dfrac{\mathrm{d}\Lambda}{\mathrm{d}l}=\Lambda\Theta.
 \end{equation}

Now for `$n$' dimensional spacetime manifold the RE becomes 
\begin{equation}\label{eq14*}
	\dfrac{1}{\Lambda}\dfrac{\mathrm{d}^{2}\Lambda}{\mathrm{d}l^{2}}+\left(\dfrac{1}{\Lambda}\dfrac{\mathrm{d}\Lambda}{\mathrm{d}l}\right)^{2}\left(\dfrac{1}{n}-1\right)+2\sigma^{2}+\tilde{R}=0,
	\end{equation} where $\tilde{R}=R_{ab}v^{a}v^{b}$. This second order nonlinear differential equation has a first integral of the form \begin{equation} \left(\dfrac{\mathrm{d}\Lambda}{\mathrm{d}l}\right)^{2}=F(\Lambda)\Lambda^{-2\left(\dfrac{1}{n}-1\right)}+q_{0}\Lambda^{-2\left(\dfrac{1}{n}-1\right)},
\end{equation}
 with $q_{0}$, an integration constant and
  \begin{equation}
 	F(\Lambda)=\int(\tilde{R}+2\sigma^{2})\Lambda^{2\left(\frac{1}{n}-1\right)}d\Lambda.
 	\end{equation}

 Here, it is assumed that $\tilde{R}+2\sigma^{2}$ is a function of $\Lambda$. Equivalently, the expression for expansion scalar becomes
   \begin{equation}
   	\Theta^{2}=q^{-\frac{1}{n}}\left[\tilde{F}(q)+q_{0}\right],
   	\end{equation}
 where $\tilde{F}(q)=\frac{1}{2}\int{(\tilde{R}+2\sigma^{2})q^{(\frac{1}{n}-\frac{3}{2})} dq}$.

  Now there is a natural question whether there exists a Lagrangian for which the Euler-Lagrange equation coincides with the RE in the form of a second order differential equation (\ref{eq14*}). For this let us start with the functional , \begin{equation}\Phi=\frac{\Lambda''}{\Lambda}+\dfrac{\Lambda'^{2}}{\Lambda^{2}}\left(\frac{1}{n}-1\right)+2\sigma^{2}+\tilde{R}.
  	\end{equation}
  Usually, a second order differential equation will be the Euler-Lagrange equation corresponding to a Lagrangian if it satisfies the Helmholtz conditions \cite{Davis:1928, Davis:1929,Douglas:1941,Casetta:1941,Crampin:2010,Nigam:2016}. Here the functional $\Phi$ given above doesn't satisfy the Helmholtz conditions. However, if we multiply $\Phi$ by $\Lambda^{\left(\frac{2}{n}-1\right)}$, then we get
\begin{equation}
\tilde{\Phi}=\Lambda^{(\frac{2}{n}-1)}\left[\frac{\Lambda''}{\Lambda}+\dfrac{\Lambda'^{2}}{\Lambda^{2}}\left(\frac{1}{n}-1\right)+\Phi_{0}(\Lambda)\right],
\end{equation}
and it can easily be verified that $\tilde{\Phi}$ satisfies all the Helmholtz conditions.
It should be noted that as $2\sigma^{2}+\tilde{R}$ is a function of $\Lambda$ alone, so we denote it by $\Phi_{0}(\Lambda)$. Then the corresponding Lagrangian is given by \cite{Choudhury:2021huy}
\begin{equation}\label{eq20}
	\mathcal{L}=\dfrac{1}{2}\Lambda^{2\left(\frac{1}{n}-1\right)}\left(\dfrac{\mathrm{d}\Lambda}{\mathrm{d}l}\right)^{2}-V[\Lambda],
\end{equation} and , hence the Hamiltonian is given by 
\begin{equation}\label{eq21}
	\mathcal{H}=\dfrac{1}{2}\Lambda^{2\left(\frac{1}{n}-1\right)}\left(\dfrac{\mathrm{d}\Lambda}{\mathrm{d}l}\right)^{2}+V[\Lambda]
\end{equation} with
\begin{equation}\label{eq22}
\dfrac{\delta{V}[\Lambda]}{\delta\Lambda}=\Lambda^{\left(\dfrac{2}{n}-1\right)}(2\sigma^{2}+\tilde{R}).
\end{equation}

It is to be noted that the functional $V[\Lambda]$ is the potential corresponding to the dynamical system representing the congruence of geodesics and it is related to the typical gravity theory.

	\section{$f(R)$-gravity in inhomogeneous FLRW model: Raychaudhuri Equation}
	In $f(R)$ gravity theory the usual Einstein-Hilbert action is generalized as
		\begin{equation}
		\mathcal{A}=\dfrac{1}{2\kappa}\int\sqrt{-g}f(R)\mathrm{d}^4x+\int\mathrm{d}^4x\sqrt{-g}\mathcal{L}_{m}(g_{\mu\nu},\sigma).
	\end{equation}
	Here $\mathcal{L}_{m}$ stands for the Lagrangian of the matter field with $\sigma$ denoting the coupling between geometry $(g_{\mu\nu})$ and matter source, $f(R)$ is an arbitrary continuous function of the Ricci scalar $R$ and $\kappa=8{\pi}G=c=1$ denotes the usual gravitational coupling. Thus the field equations for $f(R)$-gravity, obtained by variation of the above action with respect to $g_{\mu\nu}$, are in the compact form as 
    \begin{equation}\label{eq24}
    	F(R)R_{\mu\nu}-\frac{1}{2}f(R)g_{\mu\nu}-\left(\nabla_\mu\nabla_\nu-g_{\mu\nu}\Box\right)F(R)=\kappa T_{\mu\nu},
    \end{equation}
with $F(R)=\dfrac{\mathrm{d}f(R)}{\mathrm{d}R}$  and \begin{equation}T_{\mu\nu}=\dfrac{2}{\sqrt{-g}}\dfrac{\partial\left(\sqrt{-g}\mathcal{L}_{m}\right)}{\partial{g}^{\mu\nu}},
\end{equation} being the stress energy tensor of the matter field. The D'Alembertian Operator $\Box$ can be written as 
$\Box=g^{\mu\nu}\nabla_{\mu}\nabla_{\nu}$.
The trace of the field equation (\ref{eq24}) takes the form
\begin{equation}\label{eq26}
	3\Box F(R)+RF(R)-2f(R)=\kappa T.
\end{equation}

Now, combination of the field equation (\ref{eq24}) with (\ref{eq26}) gives the modified Einstein field equations (after some algebraic manipulation) as \cite{Bhattacharya:2015oma}
\begin{equation}
	G_{\mu\nu}=\tilde{T}_{\mu\nu}+\dfrac{1}{F}(\nabla_{\mu}\nabla_{\nu}F-g_{\mu\nu}N),
\end{equation}
where
\begin{equation}
\tilde{T}_{\mu\nu}=\dfrac{1}{F}T_{\mu\nu},  ~~ N(t,r)=\dfrac{1}{4}(RF+\Box F+T).
\end{equation}  Now from the field equations (\ref{eq24}) and their trace equation (\ref{eq26}) one gets for a unit time-like vector $u^{\mu}$,
\begin{equation}\label{eq7}
\tilde{R}=	R_{\mu\nu}u^{\mu}u^{\nu}=\dfrac{1}{F(R)}\left[-\dfrac{1}{2}\Box F(R)+\dfrac{1}{2}({f(R)-RF(R)})+\kappa\left(T_{\mu\nu}u^{\mu}u^{\nu}+\dfrac{1}{2}T\right)+u^{\mu}u^{\nu}\nabla_{\mu}\nabla_{\nu}F(R)\right],
\end{equation}

So for CC of a congruence of time-like curves having $u^{\mu}$ as the unit tangent vector field, the r.h.s of the above equation must be positive semi-definite. In the background of inhomogeneous FLRW space-time geometry having line-element \cite{Bhattacharya:2016env}
\begin{equation}\label{eq8}
	\mathrm{d}s^2=-\mathrm{d}t^2+a^2(t)\left[\frac{\mathrm{d}r^2}{1-\frac{b(r)}{r}}+r^2\mathrm{d}\Omega_2^2\right],
\end{equation}
where $\mathrm{d}\Omega_2^2=d\theta^{2}+\sin^{2}\theta d\Phi^{2}$ is the metric on 2-sphere ($\theta$ being the polar angle),  $a(t)$ is the scale factor, $b(r)$ is an arbitrary function of $r$, the scalar curvature has the form
\begin{equation}
	R=6(\dot{H}+2H^{2})+2\dfrac{b'}{a^{2}r^{2}}.
	\end{equation}

 The field equations for $f(R)$-gravity has the explicit form
\begin{eqnarray}
	3H^2+\frac{b'(r)}{a^2r^2}&=&\frac{\rho(r,t)}{F(R)}+\frac{\rho_e(r,t)}{F(R)}\label{eq32}\\
	-\left(2\dot{H}+3H^2\right)-\frac{b(r)}{a^2r^3}&=&\frac{p_r(r,t)}{F(R)}+\frac{p_{re}(r,t)}{F(R)}\label{eq33}\\
	-\left(2\dot{H}+3H^2\right)-\frac{(b-rb')}{2a^2r^3}&=&\frac{p_t(r,t)}{F(R)}+\frac{p_{te}(r,t)}{F(R)}\label{eq34}
\end{eqnarray}
where $H=\dfrac{\dot{a}}{a}$ is the usual Hubble parameter. Here the matter is in the form of cosmic anisotropic fluid with $\rho=\rho(r,t)$, $p_{r}=p_{r}(r,t)$, $p_{t}=p_{t}(r,t)$ as the energy density, radial and transverse pressures respectively. Also the expression for the hypothetical matter components are 
\begin{equation}
	\rho_e=N+\ddot{F} , ~ p_{re}=-N-H\dot{F}+\frac{(r-b)}{a^2r}F'-\frac{(b-rb')}{2a^2r^2}F', ~p_{te}=-N-H\dot{F}+\frac{(r-b)}{a^2r^2}F',
\end{equation} 

Now the conservation relations for the anisotropic fluid can be written as :
\begin{equation}\label{eq14}
	\frac{\partial\rho}{\partial t}+(3\rho+p_r+2p_t)H=0,\mbox{~ and ~}
	\frac{\partial p_r}{\partial r}=\frac{2}{r}(p_t-p_r)
\end{equation}

Although from cosmological principle the universe is on the large scale homogeneous and isotropic and almost all models in cosmology has this property due to the elegance and simplicity of these models, still the universe is not fundamentally homogeneous on the scale of galaxies clusters and superclusters -- there is clumping of matter. Further, at the very early phase of the universe, it is very likely to have a state of much disorder (for example in emergent era/inflationary epoch). Moreover, it is possible to have apparent acceleration of the universe due to the back reaction on the metric of the local inhomogeneities \cite{Pascual-Sanchez:1999xpt,Rasanen:2006kp}. Thus it is reasonable to consider the inhomogeneous model as an alternative to dark energy. For the present $f(R)$ gravity model it is reasonable to choose 
\begin{equation}\label{eq15}
	b(r)=b_{0}\left(\dfrac{r}{r_{0}}\right)^{3}+d_{0}=\mu_{0}r^{3}+d_0 
\end{equation}so that
\begin{equation}\label{eq38}
	R=6(\dot{H}+2H^{2})+\frac{6\mu_{0}}{a^{2}}
\end{equation}
is a function of `$t$' alone. This choice of $b(r)$ includes two parameters, namely $\mu_{0}$ and $d_{0}(\neq0)$, where $d_{0}$ is identified as the inhomogeneity parameter. The motivation behind choosing a suppressed 3rd order polynomial (not any general degree polynomial or other analytic function) lies in the fact that this particular choice of $b(r)$ with $d_{0}=0$ reduces the present model to FLRW and hence successfully helps us to study an inhomogeneous model as an alternative to dark energy. Moreover with this choice of $b(r)$, the scalar curvature $R$ given by (\ref{eq38}) turns out to be homogeneous which leads to a lot of simplification in mathematical calculations. The above modified Friedmann equations (\ref{eq32})-(\ref{eq34}) with $b(r)$ from equation (\ref{eq15}) has a possible solution for the two matter components as
\begin{eqnarray}
	\rho=3H^2g(t) ~,~ \rho_e=\frac{3\mu_0g(t)}{a^{2}}&,& p_{re}=p_{te}=-\frac{\mu_0}{a^2}+H\dot{g(t)} , \nonumber\\ p_r=\psi(t)\left[-\left(2\dot{H}+3H^2\right)-\frac{d_0}{a^2r^3}\right]-H\dot{g}(t) &,&		p_t=g(t)\left[-\left(2\dot{H}+3H^2\right)+\frac{d_0}{2a^2r^3}\right]-H\dot{g}(t)
\end{eqnarray}

The above choice shows that the usual matter component is inhomogeneous and anisotropic in nature while the hypothetical curvature fluid is both homogeneous and isotropic in nature with $\omega_{e}=\dfrac{1}{3}\left(1+\dfrac{Ha^{2}\dot{g(t)}}{\mu_{0}}\right)$ as the expression for state parameter. Further, due to the inhomogeneity (i.e $d_{0}\neq0$) the equation of state parameters for the normal fluids i.e $\omega_{r}=\dfrac{p_{r}}{\rho}$ and $\omega_t=\dfrac{p_{t}}{\rho}$ are related linearly as 
\begin{equation}\label{eq40}
	\omega_t-\omega_r=\frac{d_0}{2a^2H^2r^3}
\end{equation}

\begin{figure}[h!]
	\begin{minipage}{0.3\textwidth}
		\centering\includegraphics[height=5cm,width=5cm]{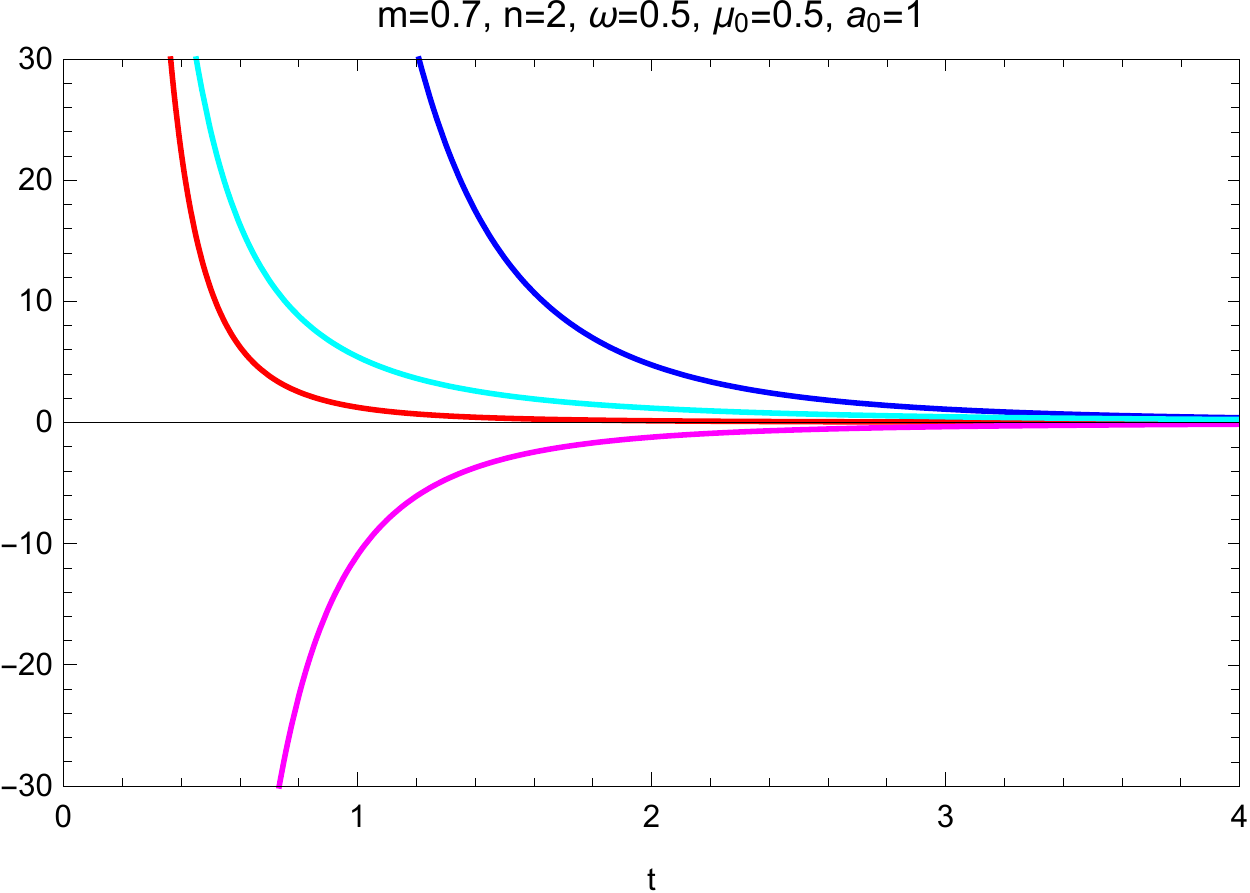}
	\end{minipage}~~~~~~~
	\begin{minipage}{0.3\textwidth}
		\centering\includegraphics[height=5cm,width=5cm]{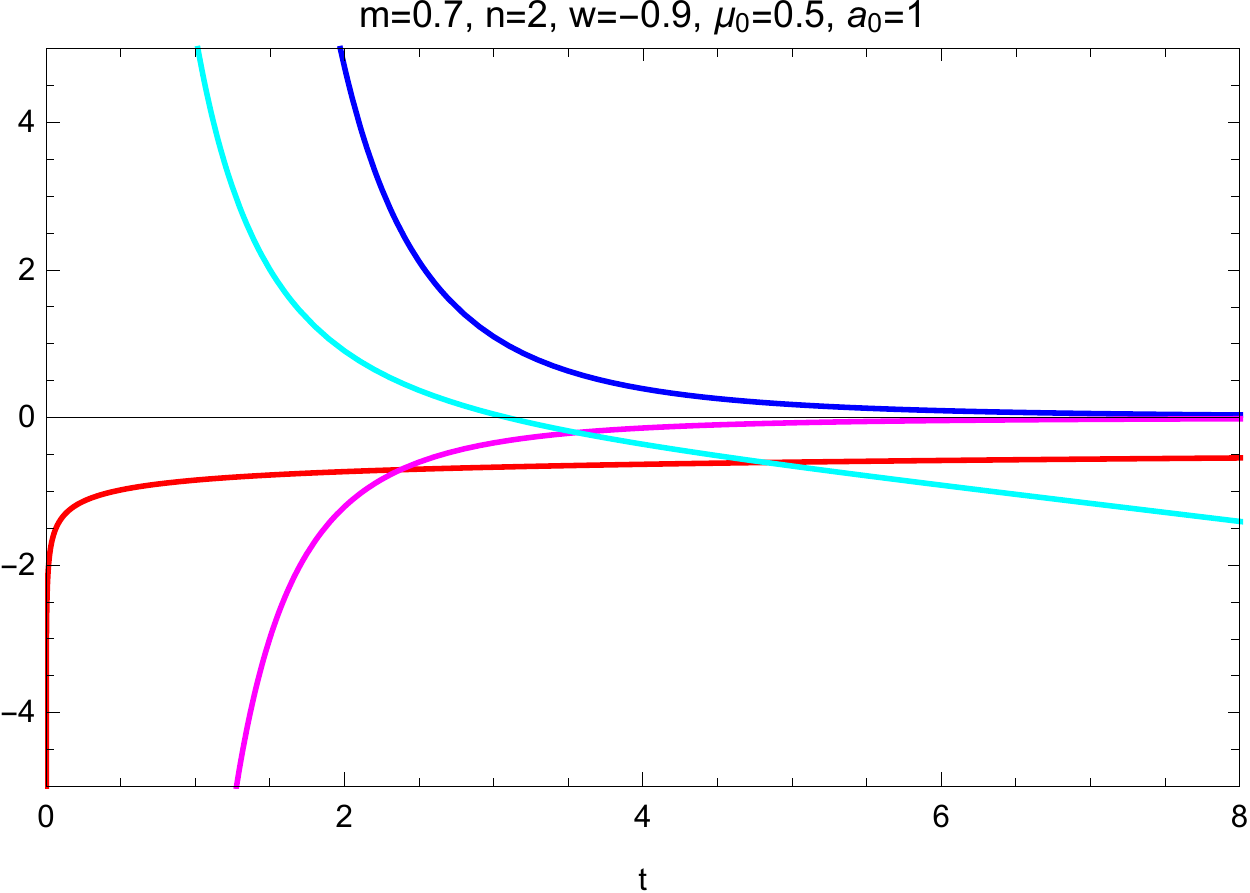}
	\end{minipage}\hfill
	\begin{minipage}{0.3\textwidth}
		\centering\includegraphics[height=5cm,width=5cm]{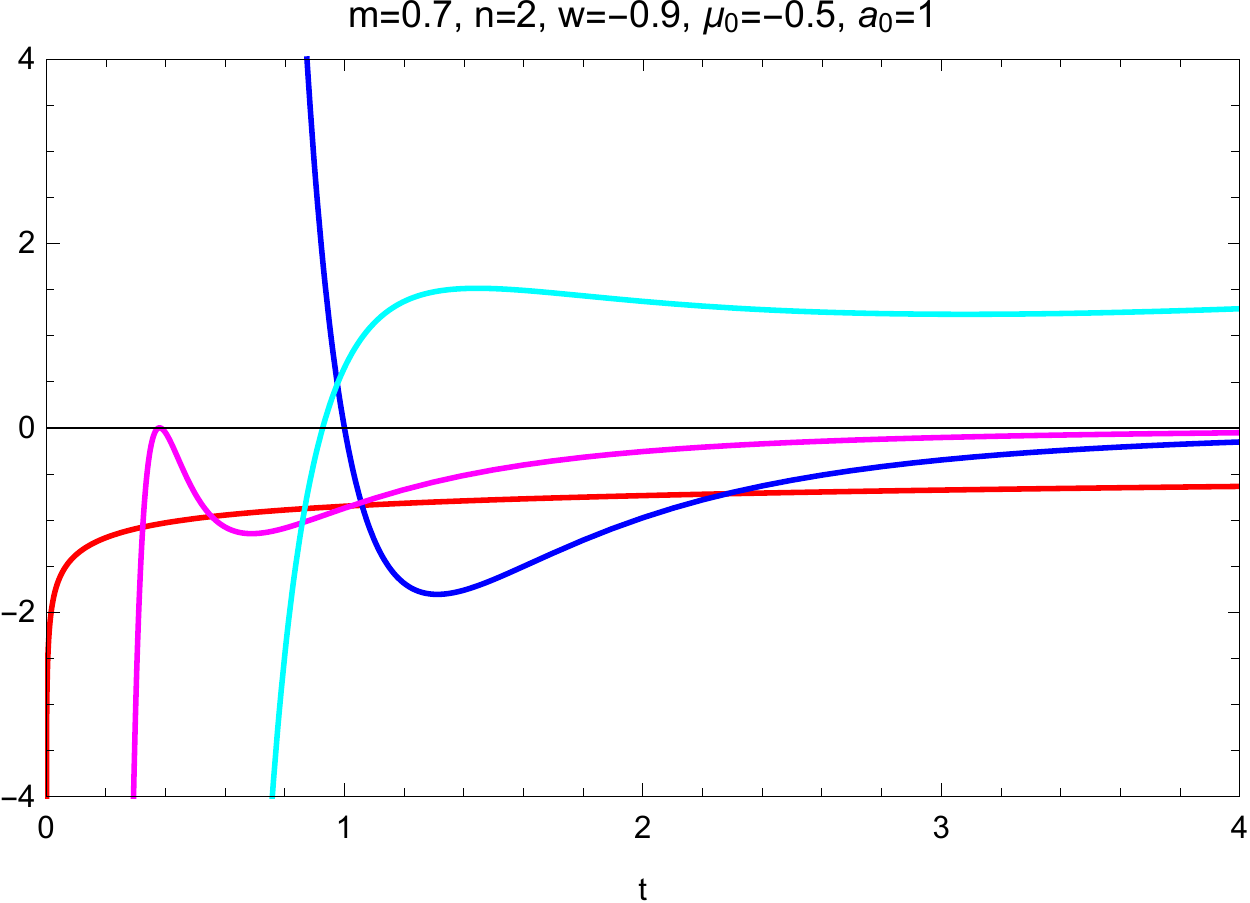}
	\end{minipage}
\begin{minipage}{0.3\textwidth}
	\centering\includegraphics[height=5cm,width=5cm]{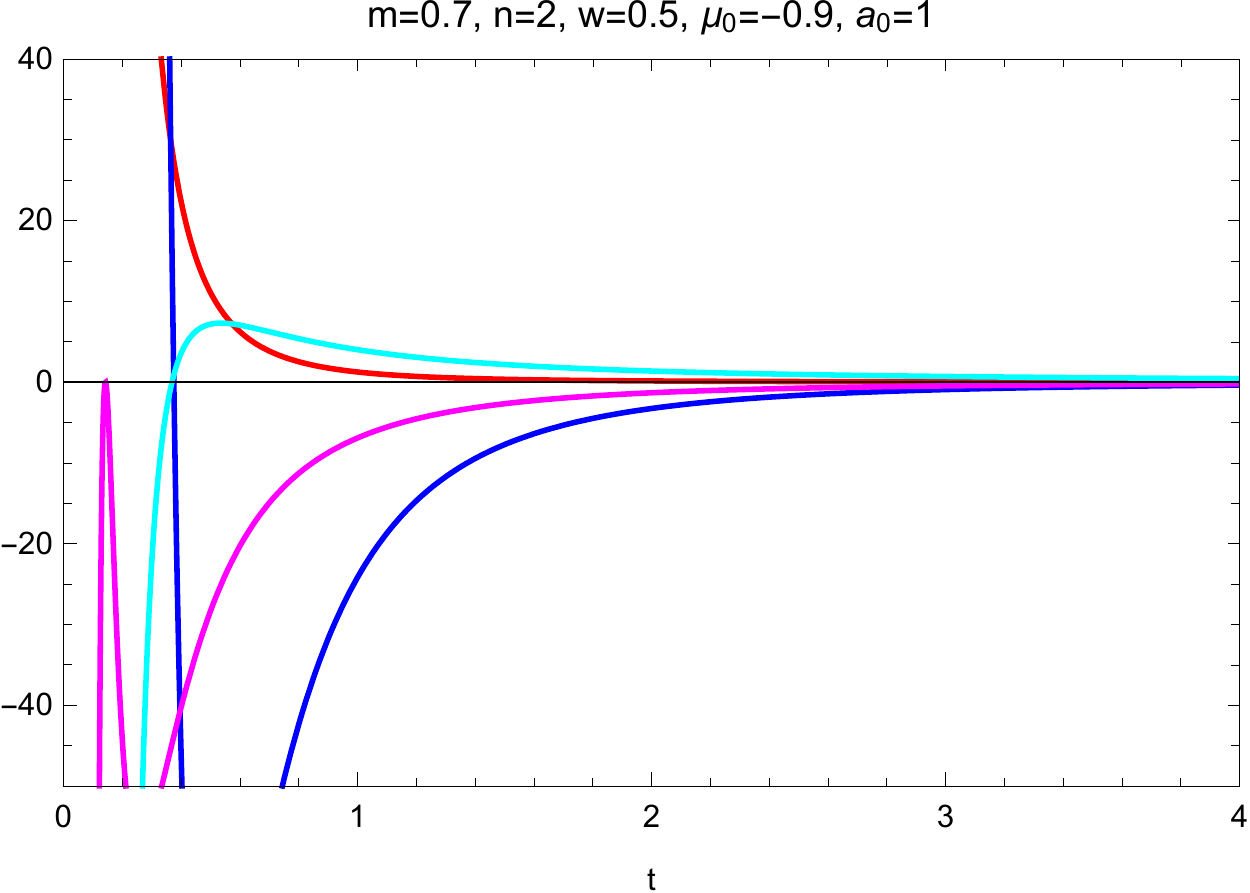}
\end{minipage}~~~~~~~
\begin{minipage}{0.3\textwidth}
	\centering\includegraphics[height=5cm,width=5cm]{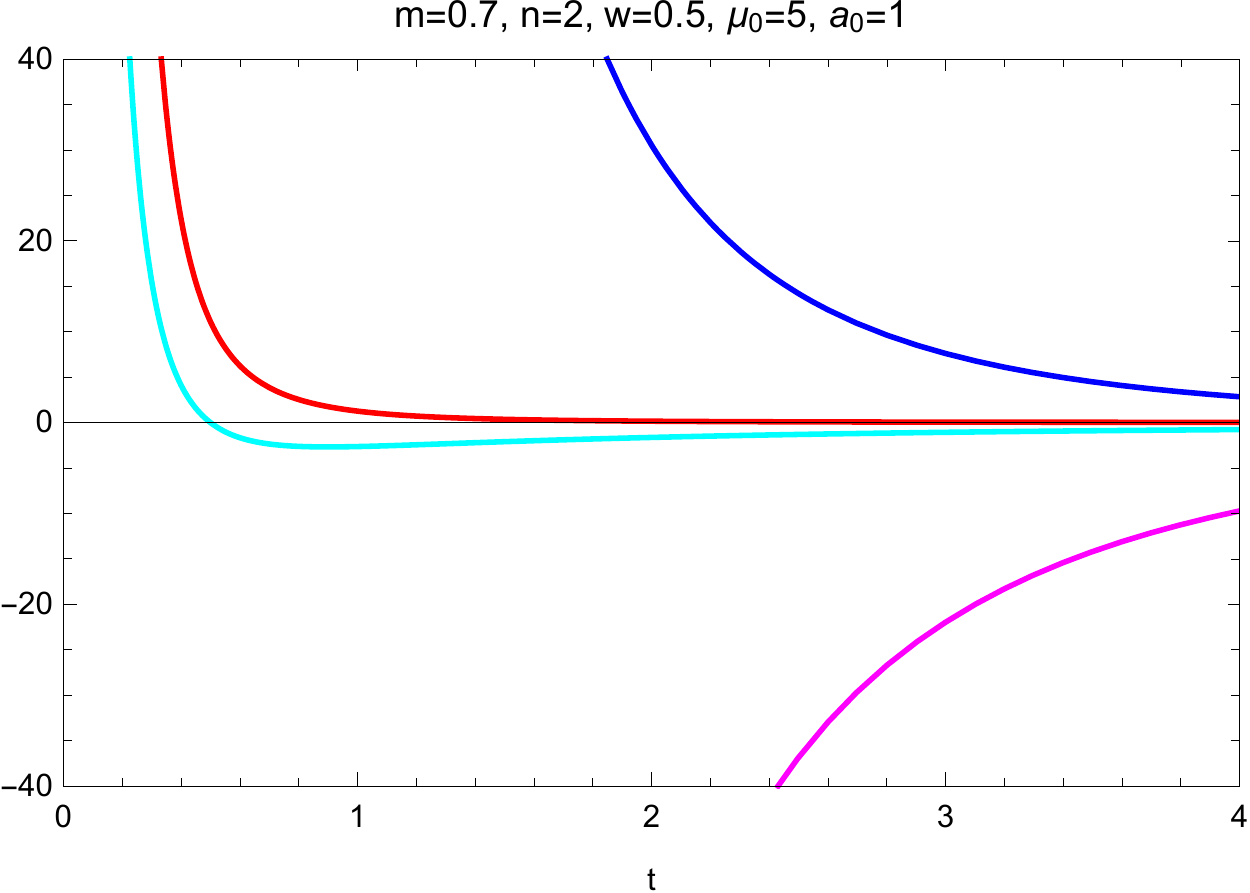}
\end{minipage}\hfill
\begin{minipage}{0.3\textwidth}
	\centering\includegraphics[height=5cm,width=5cm]{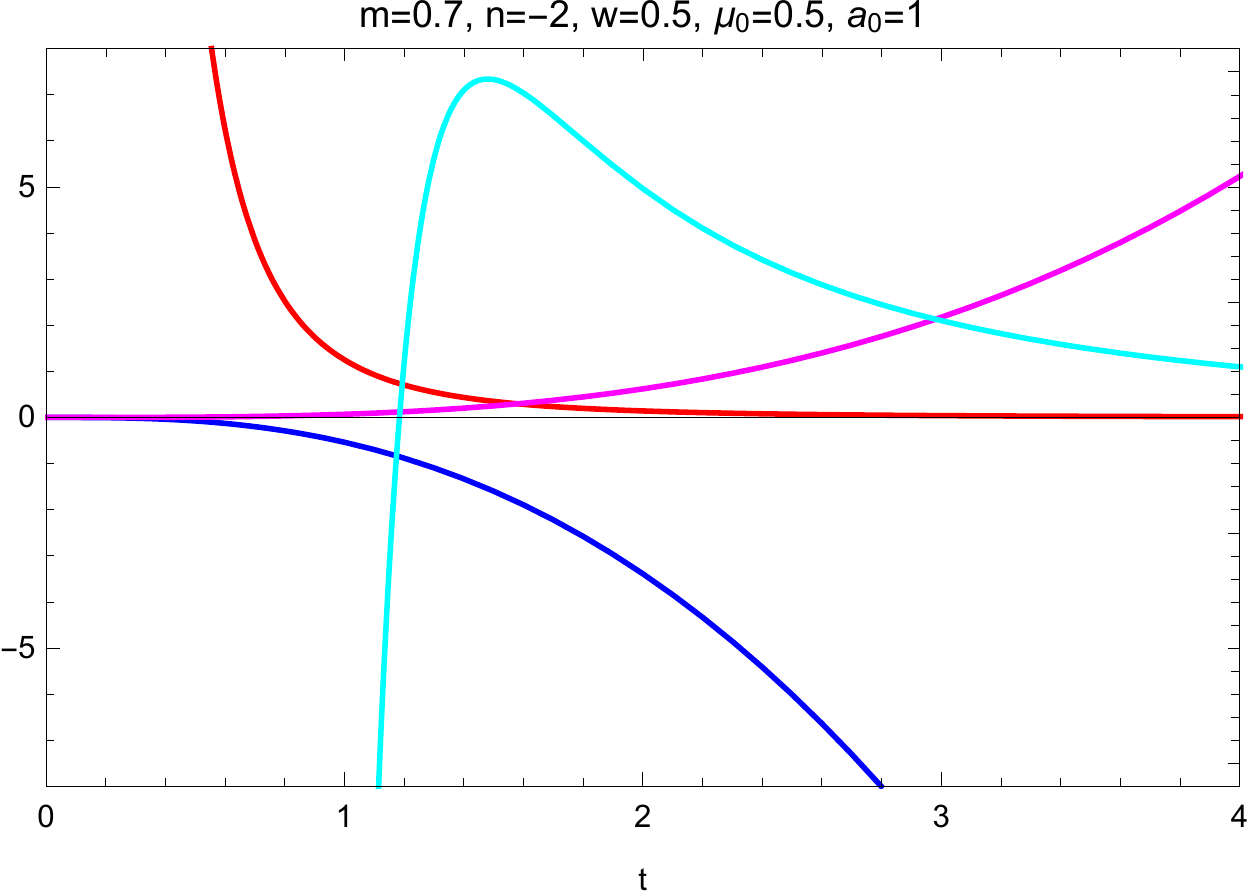}
\end{minipage}
\begin{minipage}{0.3\textwidth}
	\centering\includegraphics[height=5cm,width=5cm]{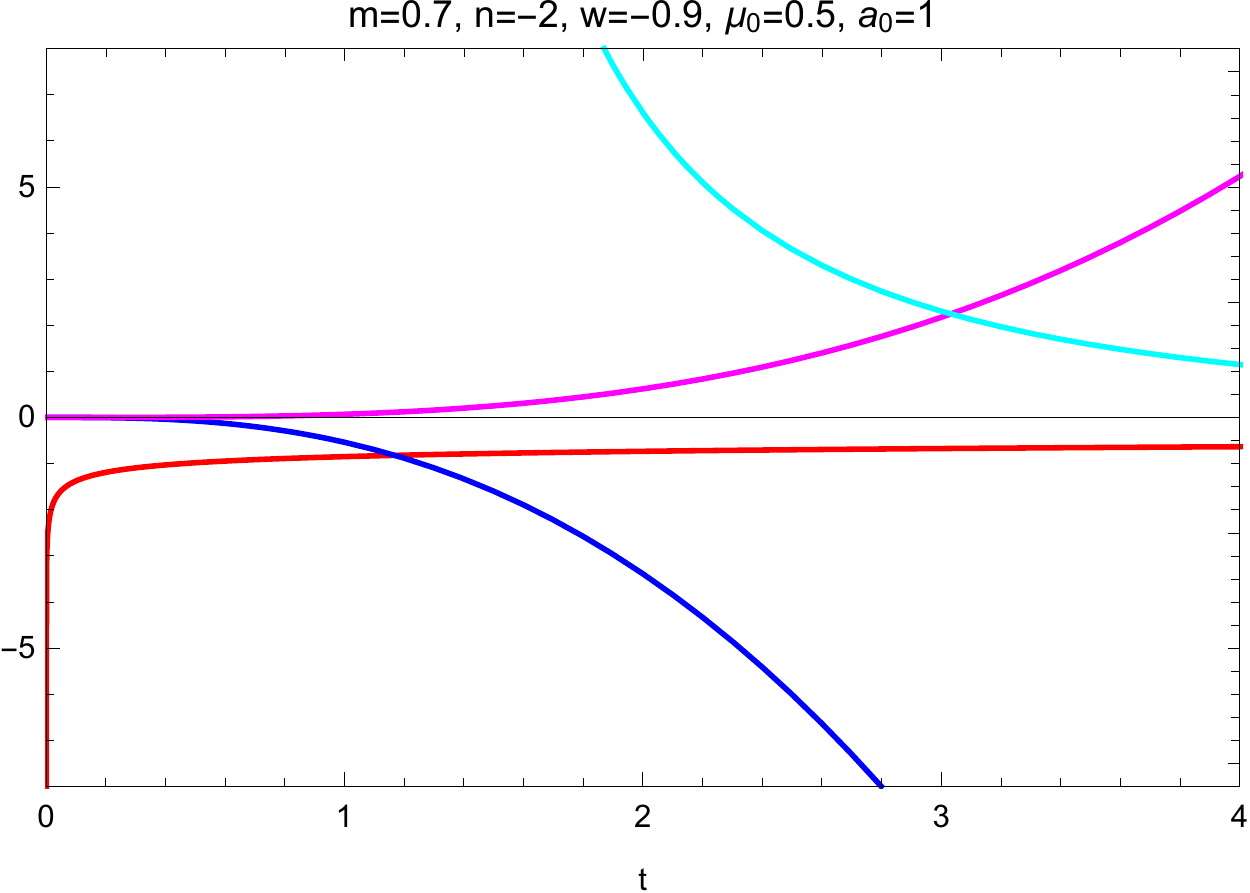}
\end{minipage}~~~~~~~
\begin{minipage}{0.3\textwidth}
	\centering\includegraphics[height=5cm,width=5cm]{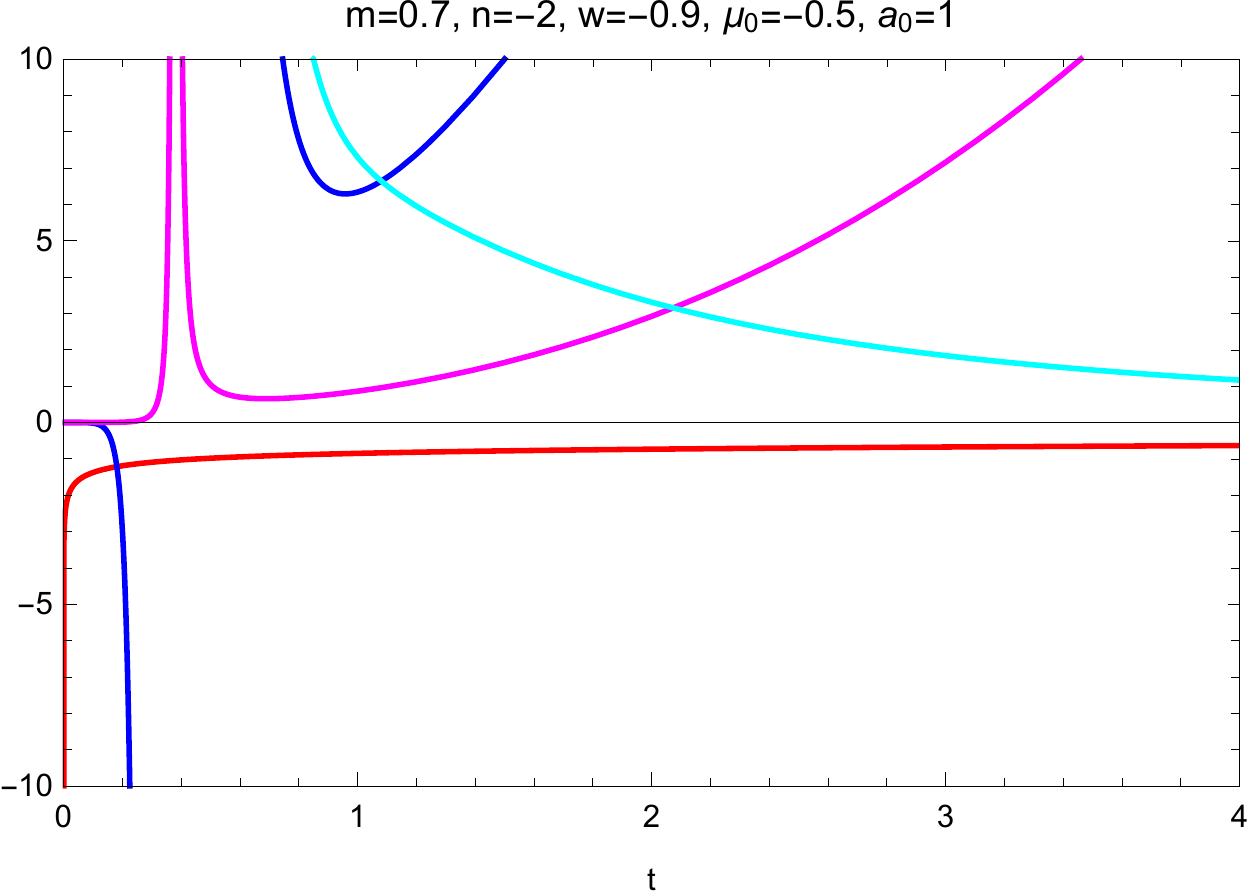}
\end{minipage}\hfill
\begin{minipage}{0.3\textwidth}
	\centering\includegraphics[height=5cm,width=5cm]{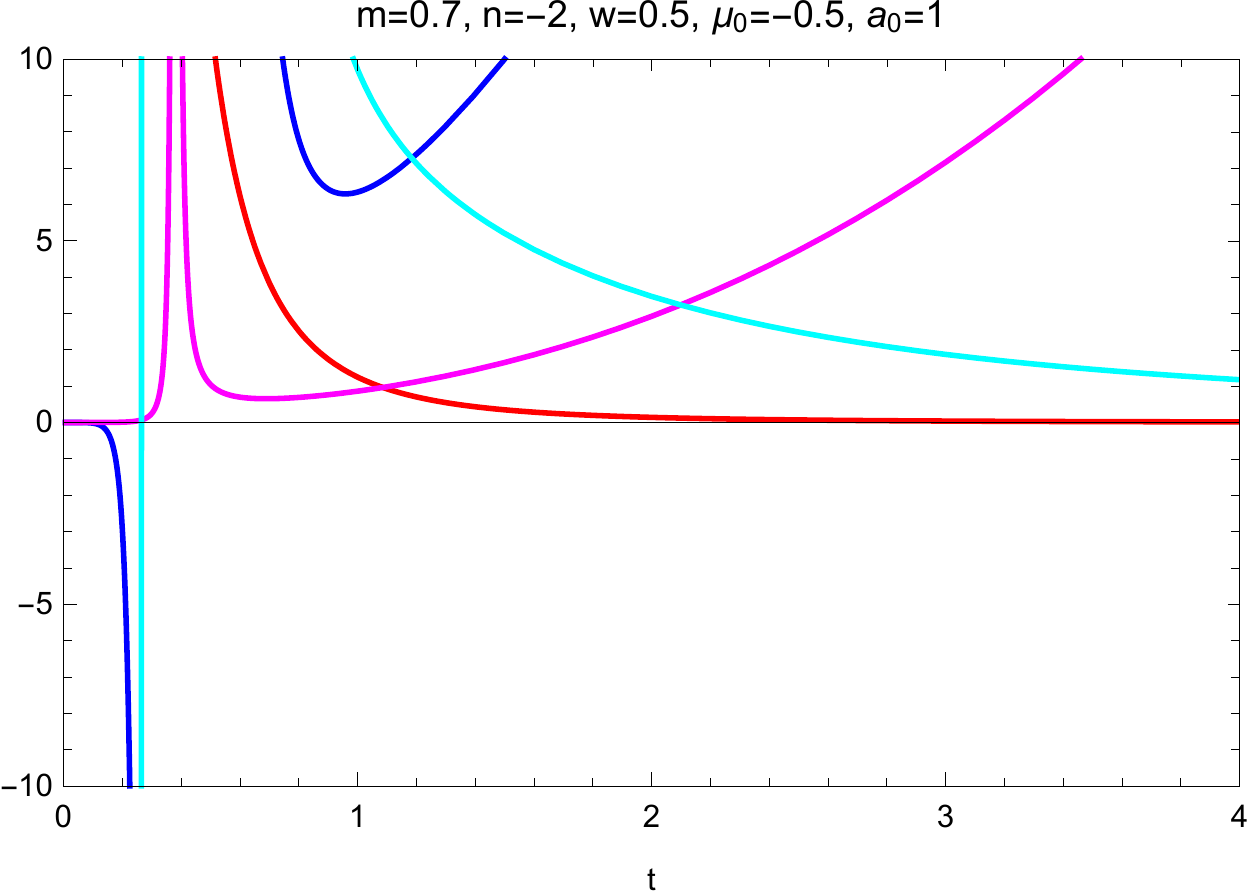}
\end{minipage}
\begin{minipage}{0.3\textwidth}
	\centering\includegraphics[height=5cm,width=5cm]{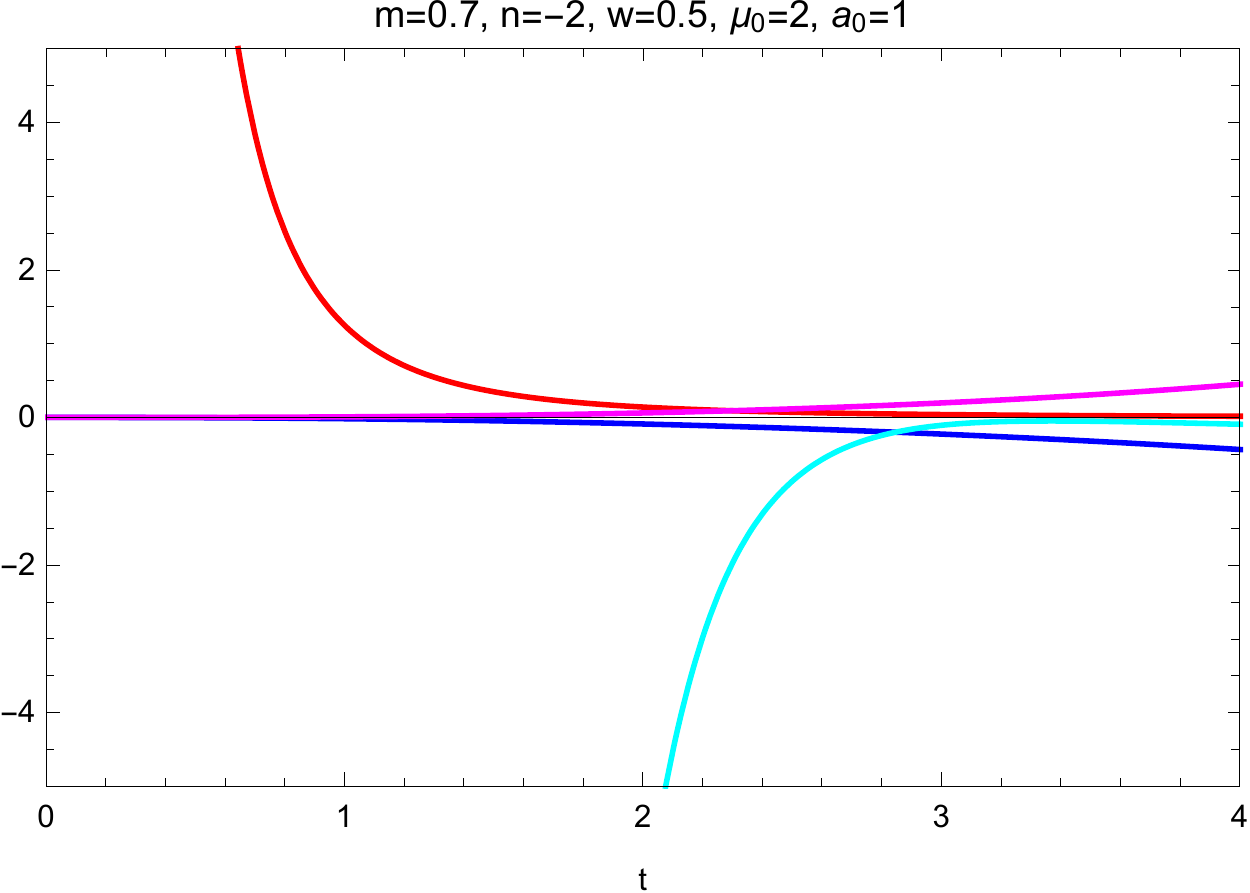}
\end{minipage}
\begin{minipage}{0.85\textwidth}\caption{Time (t) variation of different terms: $T_1$ (red), $T_2$ (blue), $T_3$ (magenta) and $\tilde{R}$= $R_{\mu\nu}u^\mu u^\nu$ (cyan) for various choices of arbitrary parameters. Here we choose $a(t)=a_{0}t^{m}$ and $f(R)=R^{n}$}\label{f1}
\end{minipage}
\end{figure}

 It may be noted that both  the conservation equations given by (\ref{eq14}) will be satisfied identically for the choice 
 \begin{equation}\label{eq19}
 	\omega_t=\frac{d_0}{6a^2H^2r^3}~ , ~ \omega_r=-\frac{d_0}{3H^2a^2r^3}
 \end{equation} 
Moreover, this choice of the state parameters results a differential equation in $\psi(t)=F(R)$ as 
\begin{equation}
	\dfrac{\dot\psi}{\psi}+2\dfrac{\dot{H}}{H}+3H=0,
\end{equation}
which has the solution 
\begin{equation}
	\psi=\frac{\psi_{0}}{(a^{3}H^{2})}
\end{equation}
This solution shows that $f(R)$ will be in the power-law form of $R$ if the power-law form of expansion of the universe is assumed. Subsequently, in the following sections we have used the power-law form of $f(R)$ particularly in graph plotting to study the CC. Also the power-law form of $f(R)$ over Einstein gravity is suitable for inflationary scenario. However, from the point of view of hypothetical curvature fluid the different components of the fluid (given by equation (\ref{eq14})) results a differential equation for $\psi$ as
\begin{equation}
	\ddot\psi-2H\dot\psi-\dfrac{2\mu_{0}}{a^{2}r^{3}}\psi=0,
\end{equation}
having solution of the form (with $\mu_{0}=0$)
\begin{equation}
	\psi(t)=\dfrac{\psi_{0}}{2}\int\dfrac{d(a^{2})}{H}
\end{equation}
Here also $f(R)$ is of the form $R^{-(n+\frac{1}{2})}$ for power law expansion of the universe.
Finally for the choice (\ref{eq19}) of the given inhomogeneous fluid after a little bit of algebra with the field equations one obtains the RE as
	\begin{equation}
			\frac{\ddot{a}}{a}=-\frac{1}{2\psi(t)}\left[\frac{\rho}{3}+H\dot{\psi}\right]
	\end{equation}
	Thus the RE is a homogeneous equation although the spacetime geometry is inhomogeneous in nature. Further, the RE does not depend on the equation of state parameters, it depends only on the energy density of the physical fluid. 
Lastly, in this section, convergence condition ($R_{\mu\nu}u^{\mu}u^{\nu} \ge0$) for a congruence of time-like curves in the present inhomogeneous space-time has been discussed. If $u^{\mu}=(1,0,0,0)$ denotes the unit time-like vector field along the congruence then from (\ref{eq7}) ,  for CC we must have
\begin{equation}
	R_{\mu\nu}u^{\mu}u^{\nu}=\dfrac{1}{F(R)}\left[T_{1}+T_{2}+T_{3}\right] \ge0
	\end{equation}
where 
\begin{eqnarray}
	T_{1}&=&\kappa\left(T_{\mu\nu} u^{\mu}u^{\nu}+\dfrac{1}{2}T\right)\\
	T_{2}&=&-\dfrac{3}{2}\Box{F(R)}=\dfrac{3}{2}u^{\mu}u^{\nu}\nabla_{\mu}\nabla_{\nu}F(R)\\
	T_{3}&=&\dfrac{1}{2}(f(R)-RF(R))
\end{eqnarray}
Note that if the matter satisfies SEC then $T_{1}\ge0$ for all time.
The time variation of $T_{1}$, $T_{2}$, $T_{3}$ and $R_{\mu\nu}u^{\mu}u^{\nu}$ has been shown graphically in FIG. \ref{f1} considering the cosmic fluid to be perfect fluid with barotropic equation of state $p=\omega\rho$. The figure shows that the CC is not universally satisfied, it depends on the choice of the parameters involved. Thus it is possible to avoid the singularity in the present model. 
\section{Geodesic congruences and Raychaudhuri Equation}
The RE for a congruence of time-like geodesics \cite{Biswas:2020hpf} with velocity vector field $v^a$ ($v^{a}v_{a}=-1$) is given by equation (\ref{eq4*}) with $A^{b}=0$. 

For the sake of simplicity, the congruence of time-like geodesics are chosen to be hyper-surface orthogonal (which by Frobenius Theorem implies zero rotation). So we have considered a metric, conformal to the original metric (\ref{eq8}) by carrying out a conformal transformation 
\begin{equation}
	dT=\dfrac{dt}{a(t)},
\end{equation}
so that the transformed metric is given by 
\begin{equation}\label{eq52*}
	ds^{2}=a^{2}(T)\left[-dT^{2}+\dfrac{dr^{2}}{1-\dfrac{b(r)}{r}}+r^{2}d\Omega_2^{2}\right]
\end{equation}

For this conformal line element, corresponding to a timelike geodesics (choosing $\theta=\dfrac{\pi}{2}$ without any loss of generality) the components of the four velocity vector field are given by 
\begin{equation}\label{eq62}
	\dot{t}=-E ,~ \dot{\phi}=\dfrac{h}{r^2}  ,~ \dot{r}=\sqrt{\left(1-\frac{b}{r}\right)\left(E^2-1-\frac{h^2}{r^2}\right)},~\dot{\theta}=0
\end{equation}
where $E$ and $h$ are identified as the conserved energy and angular momentum of the time-like particle (per unit mass). One may recall the definitions of the kinematic variables from (\ref{eq2*}) for congruence of time-like geodesics. Thus the explicit form for the kinematic variables appearing on the r.h.s of the RE are given by
\begin{eqnarray}
	\Theta^2&=&\frac{\left(1-\frac{b(r)}{r}\right)}{r^2\left(E^2-1-\frac{h^2}{r^2}\right)}\left[2(E^2-1)-\frac{h^2}{r^2}\right]^2,\\
	\sigma^2=\dfrac{1}{2}\sigma_{ab} \sigma^{ab}&=&\frac{\left(1-\frac{b(r)}{r}\right)}{3r^2\left(E^2-1-\frac{h^2}{r^2}\right)}\left[(E^2-1)^{2}+\frac{h^4}{r^4}-\frac{h^2(E^2-1)}{r^2}\right],
\end{eqnarray}
and by our construction\begin{equation}
	\omega^{2}=\dfrac{1}{2}\omega_{ab}\omega^{ab}=0
\end{equation}

\begin{figure}
		\begin{minipage}{0.3\textwidth}
		\centering\includegraphics[height=5cm,width=5cm]{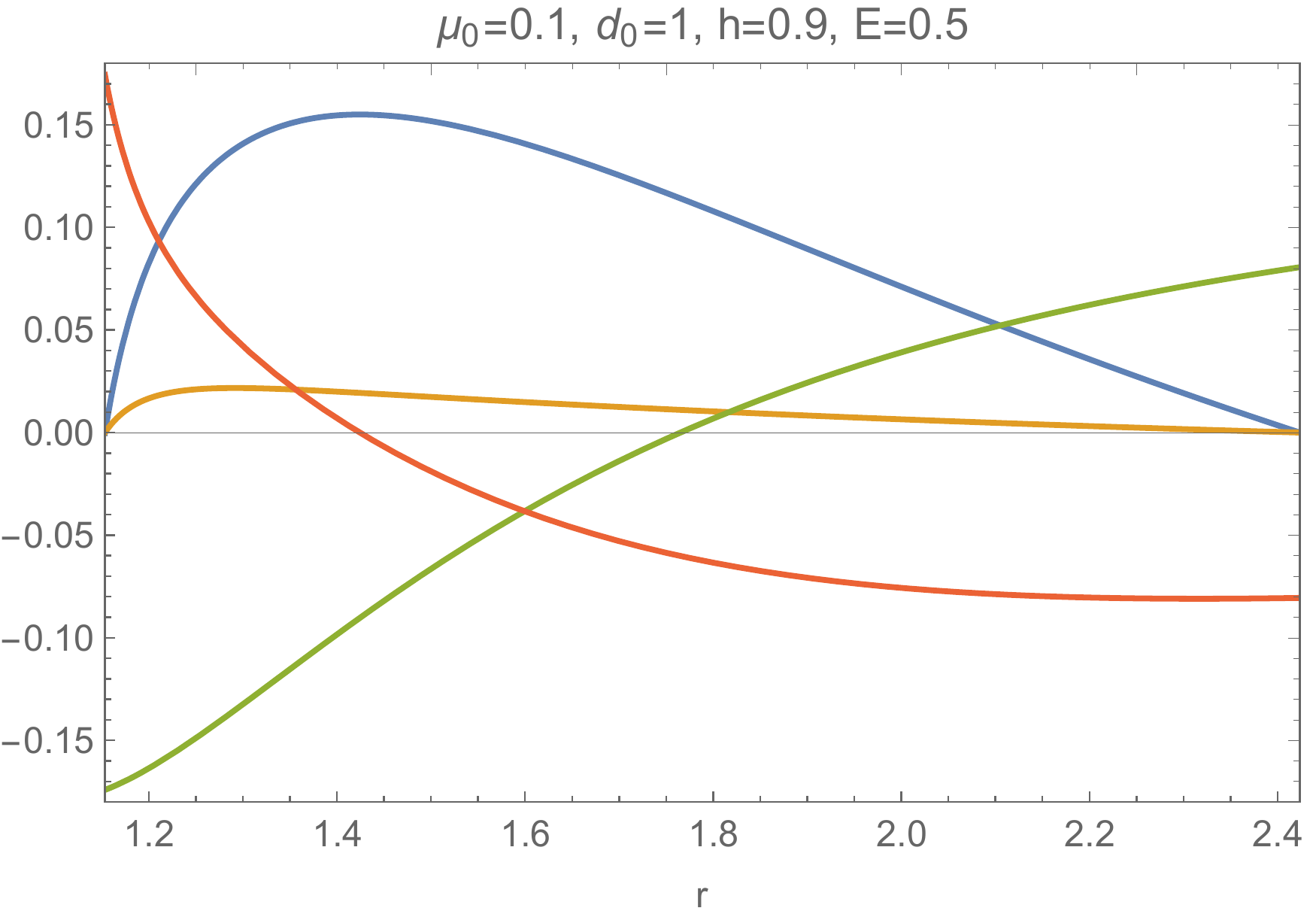}
	\end{minipage}~~~~~~~~~~
	\begin{minipage}{0.3\textwidth}
		\centering\includegraphics[height=5cm,width=5cm]{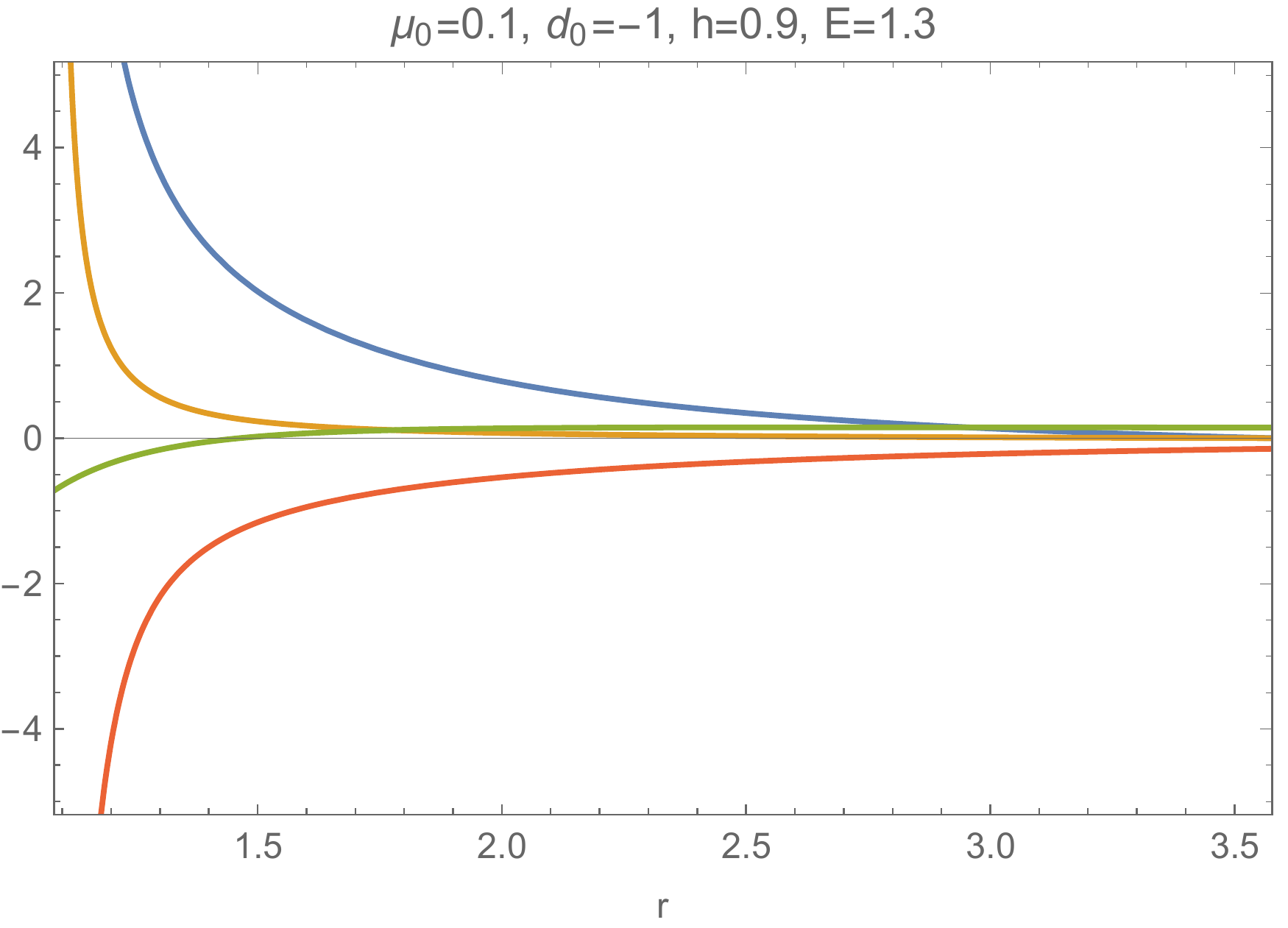}
	\end{minipage}
	\begin{minipage}{0.3\textwidth}
	\centering\includegraphics[height=5cm,width=5cm]{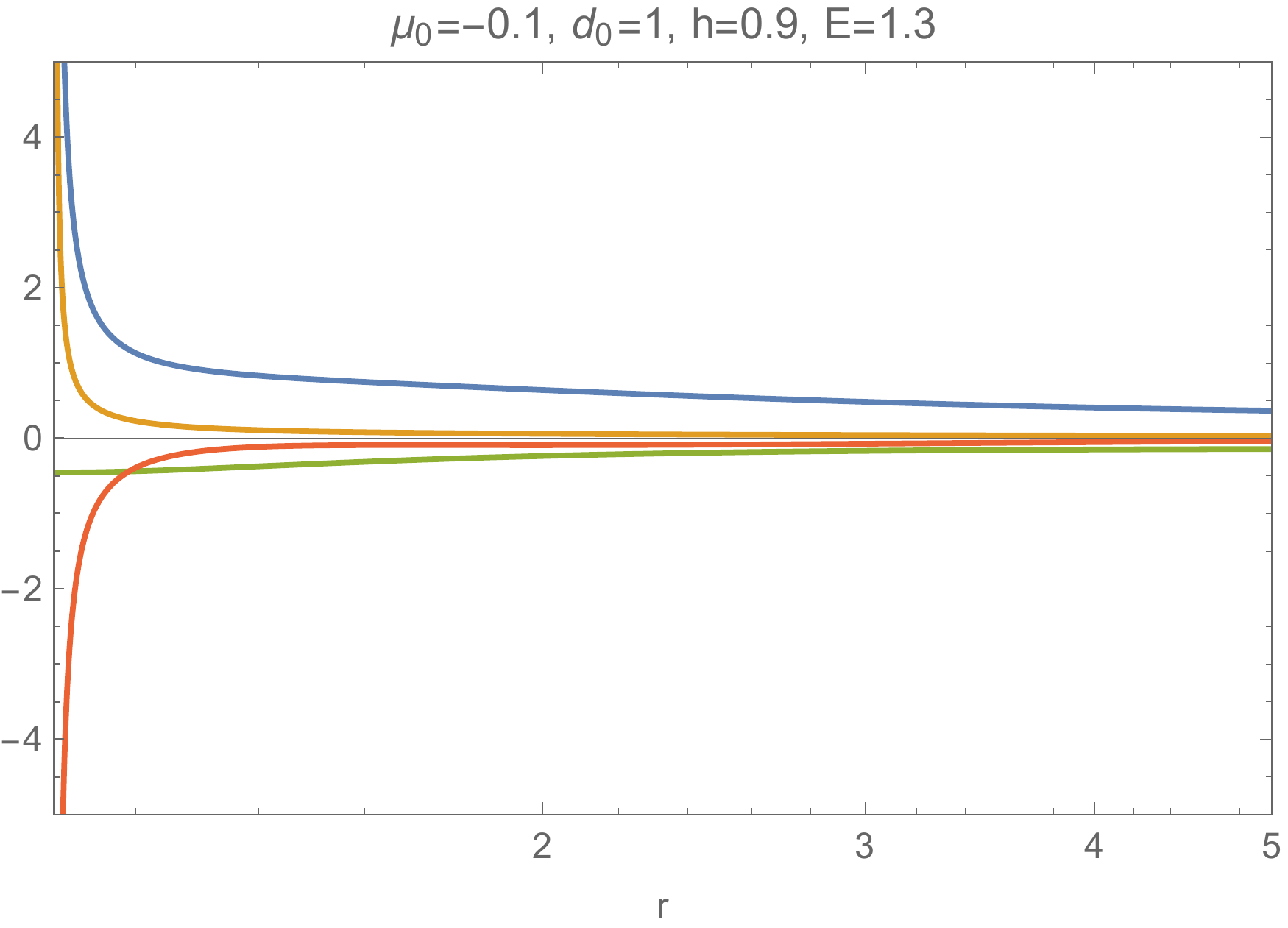}
\end{minipage}~~~~~~~~~~~
\begin{minipage}{0.3\textwidth}
	\centering\includegraphics[height=5cm,width=5cm]{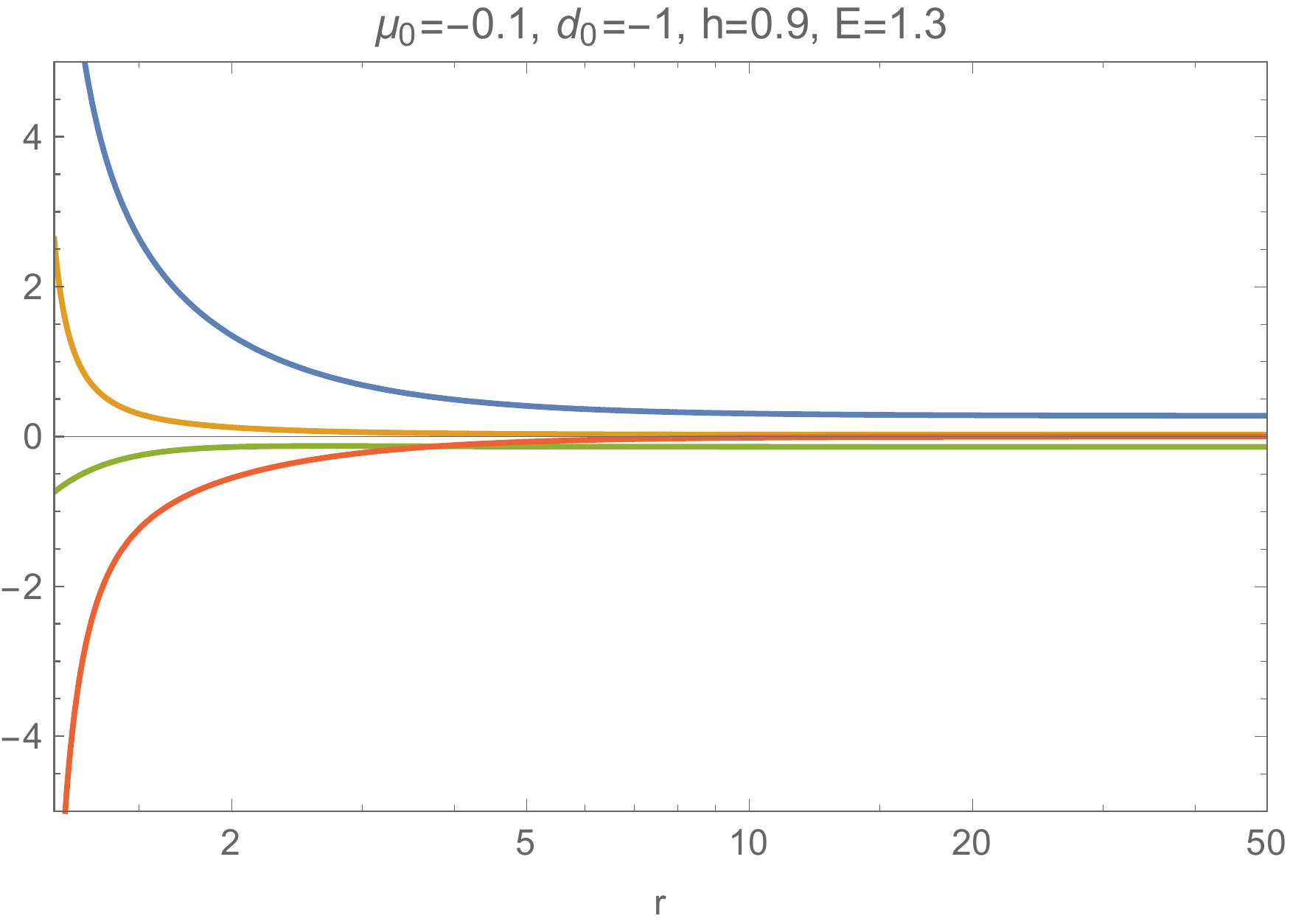}
\end{minipage}
\caption{Radial (r) variation of different kinematic parameter: $\Theta^2$ (blue), $\sigma^2$ (yellow), $R_{ab}u^a u^b$ (green) and $\dfrac{\mathrm{d}\Theta}{\mathrm{d}\tau}$ (red). From the figure it is clear that $\dfrac{\mathrm{d}\Theta}{\mathrm{d}\tau}<0$ in all cases except the first one on the top left which shows that $\dfrac{d\Theta}{d\tau}$ is positive upto some radial coordinate and then it goes negative. So $\Theta^{2}$ should be increasing upto that radial coordinate and then it is decreasing as clear from their graphs.}\label{f01}
\end{figure}

Also, the explicit expression for $R_{ab}v^a v^b$ is as follows : 
\begin{equation}
	R_{ab} v^a v^b=-\frac{h^2}{r^4}+\frac{b'(r)}{r^2}\left(E^2-1-\frac{h^2}{2r^2}\right)-\frac{b(r)}{r^3}\left(E^2-1-\frac{3h^2}{2r^2}\right)
\end{equation}
where $R_{ab}$ is the Ricci tensor projected along the congruence of geodesics and it has been evaluated from the metric (\ref{eq52*}). Now the radial variation of different kinematic parameters ($\Theta$,$\sigma$), Raychaudhuri scalar ($\tilde{R}$) and $\dfrac{d\Theta}{d\tau}$ has been studied graphically in FIG.\ref{f01} choosing parameters $\mu_{0}$, $h$ and $E$ which gives realistic cases.\\

 For the choice of $b(r)$ (given by (\ref{eq15})), the behaviour of the above kinematic variables and $R_{ab} v^{a}v^{b}$ as $r\rightarrow\infty$  are given by
\begin{eqnarray}
	\Theta^2&\xrightarrow{r\rightarrow\infty}&4\mu_0\left(1-E^2\right) \label{eq52}\\
	\sigma^2&\xrightarrow{r\rightarrow\infty}&\frac{1}{3}\mu_0\left(1-E^2\right)\label{eq53}\\
	R_{ab}v^a v^b&\xrightarrow{r\rightarrow\infty}&2\mu_0\left(E^2-1\right)\\
	\frac{\mathrm{d}\Theta}{\mathrm{d}\tau}&\xrightarrow{r\rightarrow\infty}&0\label{eq63}
\end{eqnarray}
The condition in equation (\ref{eq63}) states that the expansion leads to a constant value at infinity, and hence, the congruence will be either convergent or divergent. From the expression of $\dot{r}$ given in equation (\ref{eq62}), it can be inferred that for realistic $\dot{r}$ and $r$, one must have ($E^{2}-1)>0$, since $\left(1-\dfrac{b(r)}{r}\right)>0 $. Further it is to be noted that for real expansion scalar $\Theta$ and shear $\sigma$ at infinity $\mu_{0}$ should be negative i.e it dictates open geometry. On the other hand,  if $0<\mu_{0}<\dfrac{4}{27d_{0}^{2}}$ then geodesics are confined within a bounded region. 

\begin{figure}[h!]
		\begin{minipage}{0.3\textwidth}
		\centering\includegraphics[height=5cm,width=5cm]{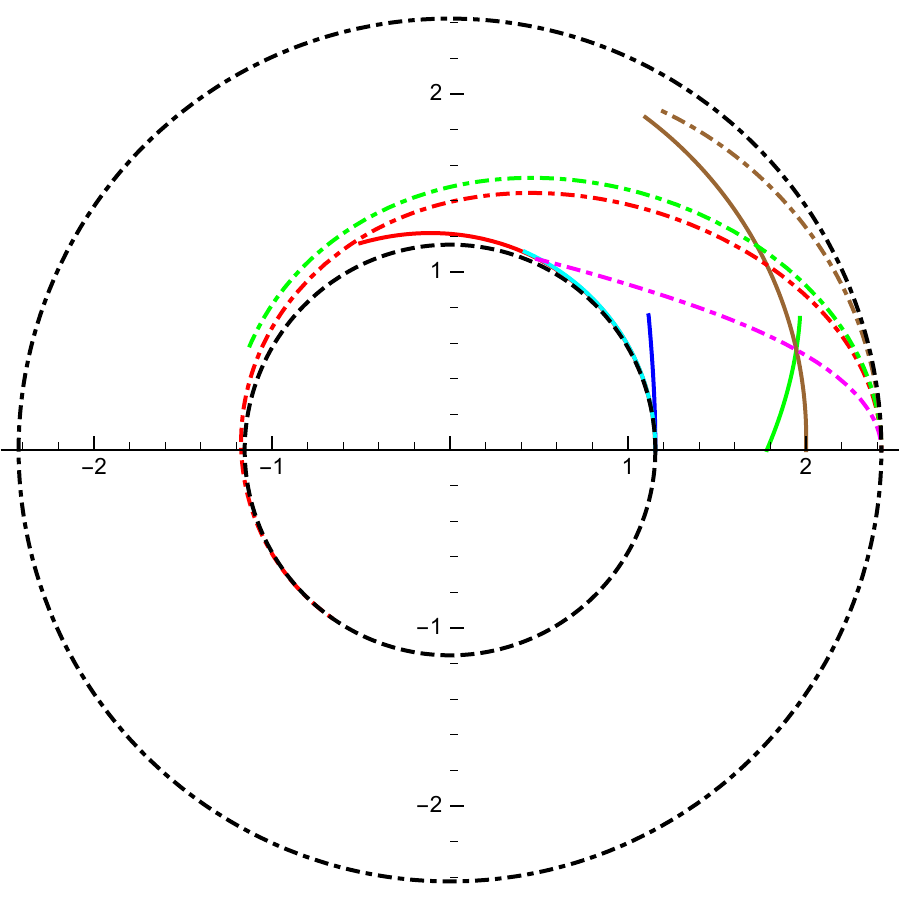}
	\end{minipage}~~~~~~~
	\begin{minipage}{0.3\textwidth}
		\centering\includegraphics[height=5cm,width=5cm]{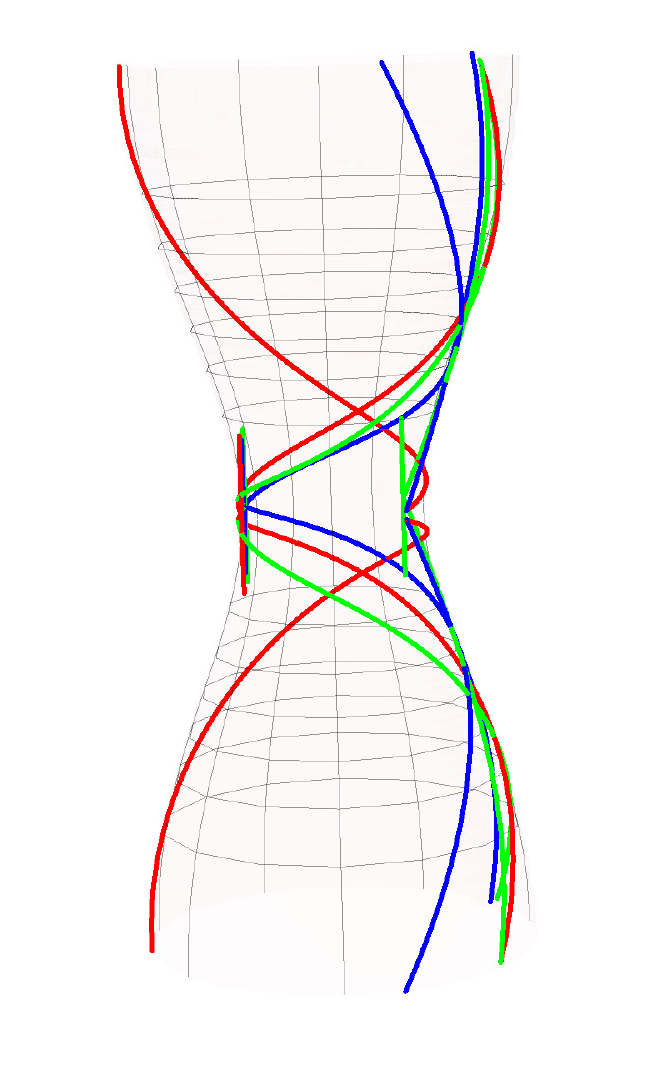}
	\end{minipage}
\caption{[ Polar plot for the congruence of geodesics for the choice of $b(r)=\mu_0r^3+d_0$ with $\mu_0=0.1$ and $d_0=1$ (left) for different initial values, namely $r_{0}=1.15348$ ( Red ,Blue, Cyan) ; $r_{0}=1.5$ ( Green , Magenta); $r_{0}=2$ ( Brown, Green(dot-dashed)); $r_{0}=2.42361$ (Red,magenta, Brown dot-dashed). Also the solid/dot-dashed graphs are drawn considering $\dfrac{\mathrm{d}r}{\mathrm{d}\phi}>0/<0$. In all cases $\phi$ has been chosen within ranges which are a proper subset of $(0,2\pi)$. For the 3D embedding diagram (right) vertical axis is z and the horizontal plane is $r-\phi$. ]}\label{f2}
\end{figure}

\begin{figure}[h!]
\centering			\includegraphics[scale=0.5]{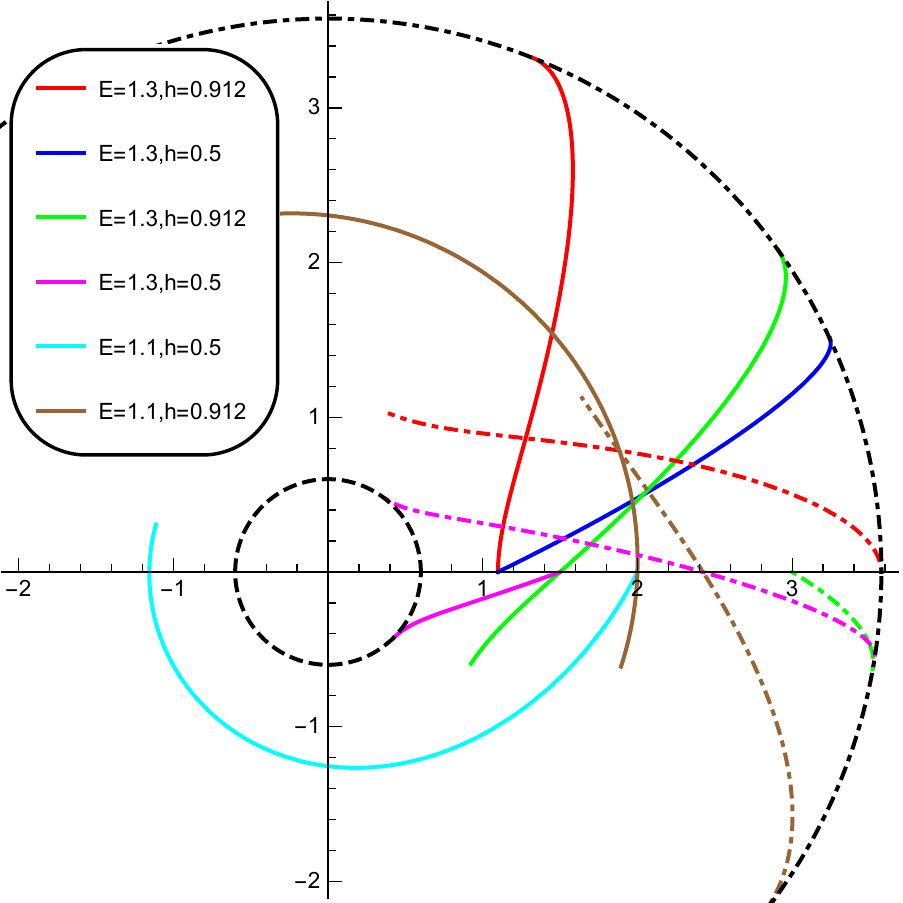}

	\caption{[Polar plot for the congruence of geodesics for the choice of $b(r)=\mu_0r^3+d_0$ with $\mu_0=0.1$ and $d_0=-1$ for different initial values,namely $r_{0}=1.098$ (Red,Blue); $r_{0}=1.5$ (Green,Magenta); $r_{0}=2$ (Cyan,Brown); $r_{0}=3.577$ (Red dot-dashed);  $r_{0}=3$ (Green dot-dashed); $r_{0}=2.42361$ (Magenta and Brown dot-dashed) $\phi \in (0,2\pi)$. Solid/ dot-dashed graphs are drawn considering $\dfrac{\mathrm{d}r}{\mathrm{d}\phi}>0/<0$. In all cases $\phi$ has been chosen within ranges which are a proper subset of $(0,2\pi)$.]}\label{f3}
\end{figure}

\begin{figure}
	\begin{minipage}{0.3\textwidth}
		\centering\includegraphics[height=5cm,width=5cm]{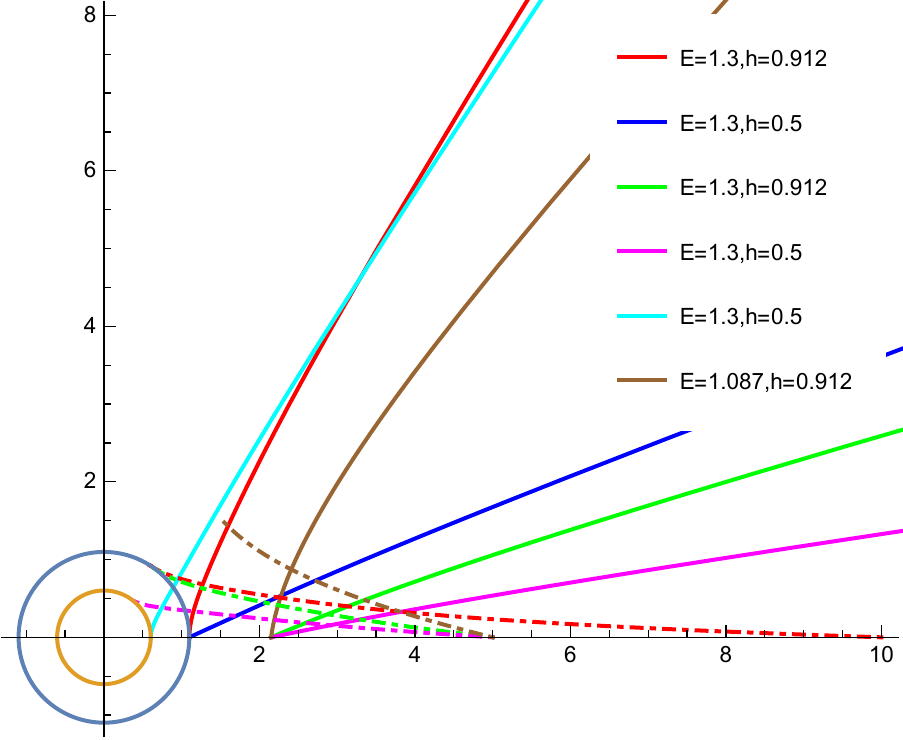}
	\end{minipage}~~~~~~~
	\begin{minipage}{0.3\textwidth}
	\centering\includegraphics[height=5cm,width=5cm]{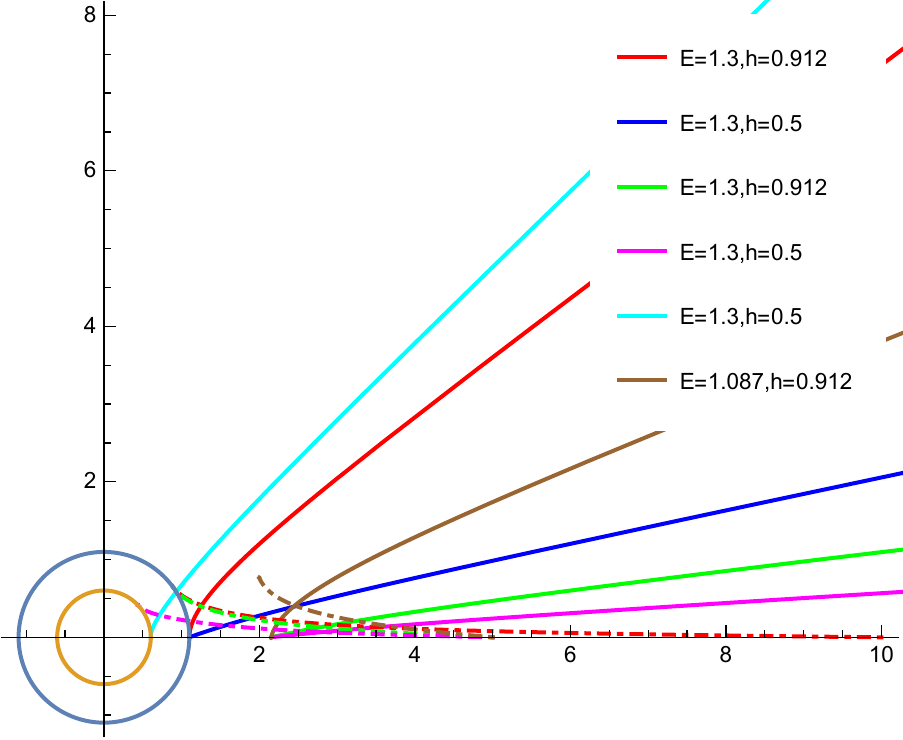}
\end{minipage}\hfill
	\begin{minipage}{0.3\textwidth}
		\centering\includegraphics[height=5cm,width=5cm]{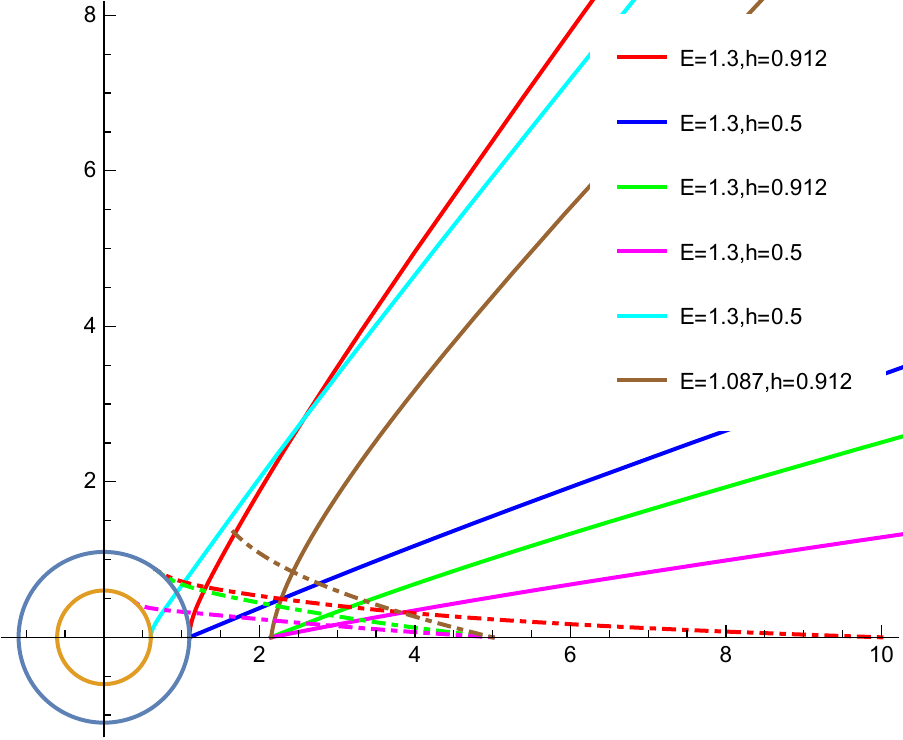}
	\end{minipage}
	\caption{[Polar plot for the congruence of geodesics for the choice of $b(r)=\mu_0r^3+d_0$ with (i) $\mu_0=-0.1$ and $d_0=-1$ (left) with initial values namely, $r_{0}=1.098$ (Red, Blue); $r_{0}=2.15$  (Green ,Magenta,Brown); $r_{0}=0.602$ (Cyan); $r_{0}=10$ (Red dot-dashed); $r_{0}=5$ (Green,Magenta,Brown dot-dashed) (ii) $\mu_0=-1.1$ and $d_0=-1$ (center) with the same initial values as (i)(left) and (iii) $\mu_0=-0.1$ and $d_0=-1.5$ (right) for the same set of initial values as in (i) and (ii). In all cases $\phi$ has been chosen within ranges which are a proper subset of $(0,2\pi)$. Solid /dot-dashed graphs are drawn considering $\dfrac{\mathrm{d}r}{\mathrm{d}\phi}>0/<0$. From the figure it can be seen that the rate of divergence (convergence) increases with decrease of $h$ and increase of $E$, $|\mu_0|$ and $|d_0|$. ] }\label{f5}
	\end{figure}

Now, to study the space-time topology and geodesic motion, let us consider the 2D hyper-surface $H: t=$constant, $\theta = \dfrac{\pi}{2}$. The geometry is characterized by \cite{Ghosh:2021dgm}
\begin{equation}\label{eq62}
	\mathrm{d}S_H^2 =\frac{\mathrm{dr^{2}}}{1- \frac{b(r)}{r}}+ r^2\mathrm{d}\Phi^2. 
\end{equation}
One may consider this 2D hyper surface $H$ embedded as rotational surface $z = z(r, \Phi)$ into the Euclidean space with metric
\begin{equation} \label{eq72}
	\mathrm{d}S_H^2 =\left[1+ \left(\frac{\mathrm{d}z}{\mathrm{d}r}\right)^2\right]\mathrm{d}r^2 + r^2\mathrm{d}\Phi^2
, \end{equation}
in cylindrical co-ordinates $(r, \phi, z)$. Thus, comparing (\ref{eq62}) and (\ref{eq72}), one may obtain the expression for embedding
function as
\begin{equation}\label{eq64}
	z(r)=\int\limits_{r_0}^r\sqrt{\frac{\frac{b(r)}{r}}{1-\frac{b(r)}{r}}}\mathrm{d}r
\end{equation}
where $r_{0}$ is a non-zero constant related to $\mu_{0}$ by the relation $\dfrac{b_{0}}{r_{0}^{3}}=\mu_{0}$ (\ref{eq15}). In comparison to FLRW metric, $\mu_{0}=\dfrac{b_{0}}{r_{0}^{3}}$ is related to the curvature scalar $\kappa$ which takes values 0, +1 and -1 for flat, closed and open model respectively.
The regions in which congruence of time-like geodesics exist  for the choice of $b(r)$ (given by (\ref{eq15})) with different (feasible) signs of $\mu_{0}$ and $d_{0}$ has been presented in Table \ref{t1} and the corresponding scenario is depicted graphically in Figures \ref{f2}-\ref{f5}. 
\begin{table}[h!]
	\caption{Regions for geodesics for different signs of $\mu_0$ and $d_0$ 	with $b(r_1)=0$, $r_\star=\dfrac{|h|}{\sqrt{E^2-1}}$ and\\ $~~~~~~~~~~~~~~~$suffix `$+$' indicates the positive root of the equation $b(r)=r$. }\label{t1}
\begin{tabular}{|c|c|c|c|c|c|}
	\hline
	Case&\multicolumn{2}{c|}{\begin{tabular}{c}Choice for\\ $\mu_0$ \& $d_0$\end{tabular}}&\begin{tabular}{c}Region for\\ $\left(1-\dfrac{b(r)}{r}\right)>0$\end{tabular}&\begin{tabular}{c}Region in which\\ geodesics exists\end{tabular}&\begin{tabular}{c}Region in which\\ geodesics are embedded\end{tabular}\\\hline
	IA &\multirow{2}{*}{$\mu_0>0$, $d_0>0$}&$\mu_0<\dfrac{4}{27d_0^2}$&$r_{+1}<r<r_{+2}$&$\max \left\{r_{+1},r_\star\right\}<r<r_{+2}$&$r_{+1}<r<r_{+2}$\\\cline{1-1}\cline{3-6}
	IB& &$\mu_0>\dfrac{4}{27d_0^2}$&not possible&not possible&not possible\\\hline
	II&\multicolumn{2}{c|}{$\mu_0>0$, $d_0<0$}&$r<r_+$&$r_\star<r<r_+$&$r_1<r<r_+$\\\hline
	III&\multicolumn{2}{c|}{$\mu_0<0$, $d_0>0$}&$r>r_+$&$r>\max \left\{r_+,r_\star\right\}$&$r_+<r<r_1$\\\hline
	IV&\multicolumn{2}{c|}{$\mu_0<0$, $d_0<0$}&$r>0$&$r>r_\star$&no embedded region\\\hline
\end{tabular}

* In case IV i.e $\mu_{0}<0$,$d_{0}<0$, the integrand becomes imaginary so there is no real expression for the embedding function $z(r)$ given by (\ref{eq64}). Hence embedding is not feasible.
\end{table}
\begin{figure}[h!]
	\begin{minipage}{0.3\textwidth}
		\centering\includegraphics[height=5cm,width=5cm]{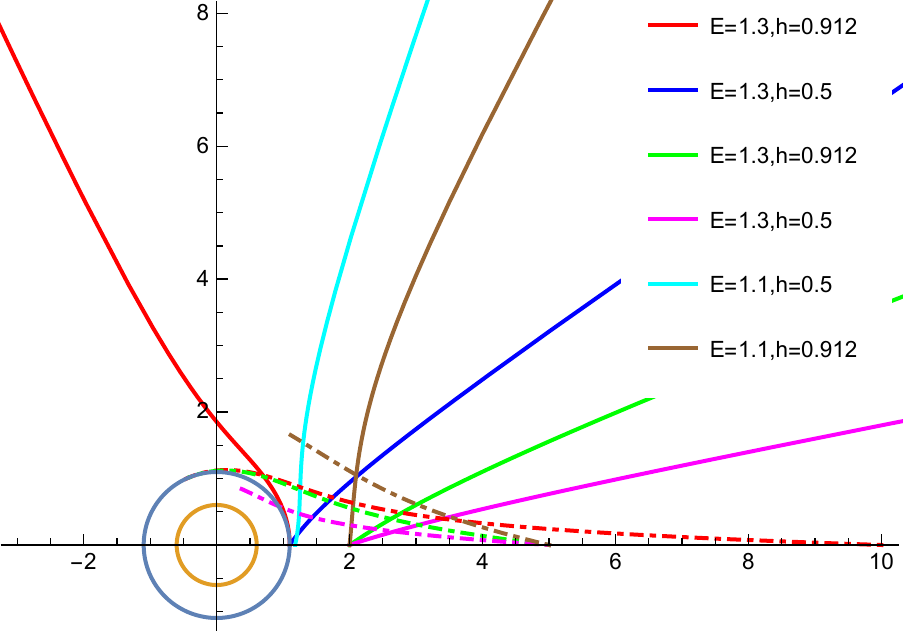}
	\end{minipage}~~~~~~~
	\begin{minipage}{0.3\textwidth}
		\centering\includegraphics[height=5cm,width=5cm]{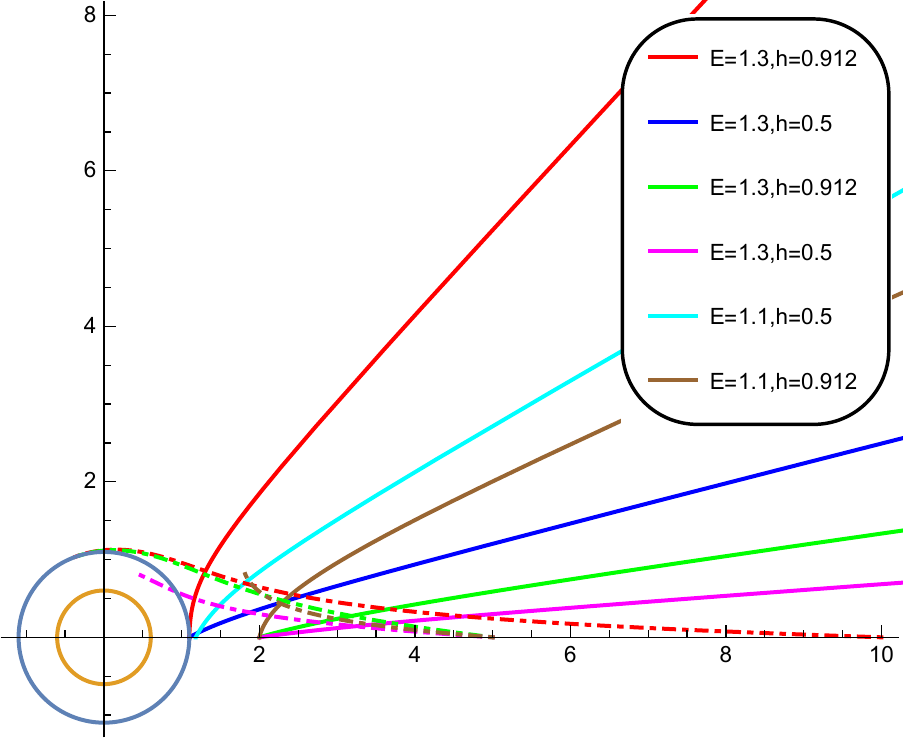}
	\end{minipage}\hfill
	\caption{[Polar plot for the congruence of geodesics for the choice of $b(r)=\mu_0r^3+d_0$ with (i) $\mu_0=-0.1$ and $d_0=1$ (left)  with initial values namely, $r_{0}=1.098$ (Red, Blue); $r_{0}=2$ (Green, Magenta, Brown);  $r_{0}=1.18$ (Cyan); $r_{0}=10$ (Red dot-dashed); $r_{0}=5$ (Green,Magenta,Brown dot-dashed). (ii) $\mu_0=-0.1$ and $d_0=1.5$ (right) for the initial values same as (i)(left). In all cases $\phi$ has been chosen within ranges which are a proper subset of $(0,2\pi)$. Solid/ dot-dashed graphs are drawn considering $\dfrac{\mathrm{d}r}{\mathrm{d}\phi}>0/<0$. From the figure it can be seen that the rate of divergence (convergence) increases with decrease of $h$ and increase of $E$, $|\mu_0|$ and $|d_0|$.]  }
\end{figure}
\section{Quantization Scheme}
We shall now describe canonical quantization of the present system described by the point-like Lagrangian (\ref{eq20}). The momentum conjugate to the configuration variable $\Lambda$ is given by
\begin{equation}
	\Pi_{\Lambda}=\dfrac{\partial\mathcal{L}}{\partial\Lambda^{'}}=\Lambda^{2\left(\frac{1}{n}-1\right)}\Lambda^{'}
\end{equation} 
So , the Hamiltonian of the system has the expression given by (\ref{eq21}). One can easily verify that the Hamilton's equation of motion will give the RE and the definition of the momentum.\\In canonical quantization , the (canonically conjugate) dynamical variables $\Lambda$ and $\Pi_{\Lambda}$ are considered as operators acting on the state vector $\Psi(\Lambda,\lambda)$ of the present geometric flow . In the $\Lambda$ representation, the operators take form 
\begin{equation}
	\tilde{\Lambda}\longrightarrow{\Lambda}
	\end{equation} and\begin{equation}
	\tilde{\Pi}_{\Lambda}\longrightarrow{-i\hbar\dfrac{\partial}{\partial{\Lambda}}}
\end{equation} with commutation relation
\begin{equation}
	[\tilde{\Lambda}  , \tilde{\Pi}_{\Lambda}]=i\hbar
\end{equation}
So the evolution of the state vector is given by
\begin{equation}\label{eq71}
	i\hbar\dfrac{\partial\Psi}{\partial\lambda}=\tilde{H}\Psi
\end{equation} with ,
\begin{equation}
	\tilde{H}=-\dfrac{\hbar^{2}}{2}\Lambda^{2(1-\frac{1}{n})}\dfrac{\partial^{2}}{\partial\Lambda^{2}}+V[\Lambda]
\end{equation} being , the operator version of the Hamiltonian. This evolution equation can be termed as the quantized RE. Further, the evolution equation indicates the motion of the quantized congruence of time-like geodesics.\\
On the other hand , in context of cosmology , there is Hamiltonian constraint and the operator version of it acting on the wave function of the universe which is termed as Wheeler-Dewitt (WD) equation \cite{Wheeler:1968iap} i.e
\begin{equation}
	\tilde{\mathcal{H}}\Psi=0.
\end{equation} There is a problem of non-unitary evolution \cite{Pinto-Neto:2013toa,Halliwell:1989myn,Pal:2014xsa,Alvarenga:2003kx,Pal:2014dya} of the system which can be resolved by the appropriate choice of operator ordering in the first term of the Hamiltonian. Note that the operator version (\ref{eq71}) is a symmetric operator with norm
\begin{equation}
	\|\Psi\|=\int_{0}^{\infty}d\Lambda \Lambda^{2\left(\frac{1}{n}-1\right)}\Psi^{*}\Psi ,
\end{equation}
but it is not a self adjoint operator. However , one can extend it as a self adjoint operator with the following operator ordering \cite{Pal:2016ysz}
\begin{equation}
	\tilde{H}=-\dfrac{\hbar^{2}}{2}\Lambda^{(1-\frac{1}{n})}\dfrac{\partial}{\partial\Lambda}\Lambda^{(1-\frac{1}{n})}\dfrac{\partial}{\partial\Lambda}+V[\Lambda]
\end{equation}
A change of the minisuperspace variable namely 
\begin{equation}
	u=n\Lambda^{\frac{1}{n}}
	\end{equation}
simplifies the WD equation as 
\begin{equation}
	\left[\dfrac{-\hbar^{2}}{2}\dfrac{d^{2}}{du^{2}}+V(u)\right]\Psi(u)=0
\end{equation}
with $V(u)$, given by solution of equation (\ref{eq22}) and symmetric norm given by
\begin{equation}
\|\Psi\|=\int_{0}^{\infty}du\Psi^{*}\Psi
\end{equation}	
This self adjoint extension of the Hamiltonian operator simplifies the WD equation to one dimensional Helmholtz equation with varying wave number and the wave function $\Psi$ may be interpreted as the amplitude of the propagation of the congruence of geodesics .
\section{Brief Discussion and Conclusion}

An extensive analysis of the Raychaudhuri equation has been done in the present work for $f(R)$ modified gravity theory in the background of inhomogeneous FLRW space-time. By suitable transformation, this first order nonlinear ordinary differential equation can be converted to a second order linear differential equation analogous to the evolution equation for simple harmonic oscillator (with varying frequency). Further, this 2nd order differential equation written in terms of the three metric has a first integral to have an explicit form of the expansion scalar. Also it is possible to have a Lagrangian from which this second order evolution equation can be derived as Euler-Lagrange equation. Though the present model is inhomogeneous but still the RE so constructed turns out to be a homogeneous differential equation. For congruence of time-like geodesic , convergence  conditions have been analyzed graphically and one may conclude that singularity may be avoided for specific choices for the parameters involved. Finally, congruence of time-like geodesics are studied for the present model , choosing the corresponding conformal metric for simplicity. Different kinematic parameters (involved in the RE) are evaluated and their asymptotic behaviours are examined at infinity. It is found that both bounded and unbounded geodesics are possible for different signs of the parameters involved in the metric function $b(r)$. Also, all possible types of geodesics are shown graphically. Finally, a canonical quantization scheme has been developed from the Hamiltonian of the system. The evolution equation is termed as quantum Raychaudhuri equation. By choosing suitable operator ordering it is possible to have a self-adjoint WD operator and the corresponding WD equation simplifies to the 1D Helmholtz equation representing the amplitude of the congruence of geodesics in the minisuperspace.
\section*{Acknowledgements}

The authors thank the anonymous referee whose comments and suggestions improved the quality of the paper. The authors M.C. and A.B. thank University Grant Commission (UGC) for providing their respective Junior (ID: 211610035684/JOINTCSIR-UGCNETJUNE2021) and Senior (ID: 1207/CSIRNETJUNE2019) Research Fellowship. 

\end{document}